\documentclass[onecolumn]{aastex}

\usepackage{bm}
\usepackage{amsmath}
\usepackage{emulateapj5}

\newcommand{\iras}{{\sl IRAS}}
\newcommand{\obeam}{\Omega_{\rm b}}
\newcommand{\aven}{\bar{n}}
\newcommand{\lff}{{\mathcal L}_f(t)}
\newcommand{\lfg}{{\mathcal L}_g(t)}
\newcommand{\lfgn}{{\mathcal L}_{g_N}(t)}
\newcommand{\prob}[1]{\mathsf{Prob}\left\{#1\right\}}
\newcommand{\vect}[1]{{\bm #1}}
\newcommand{\expc}[1]{\mathsf{E}\left[#1\right]}
\newcommand{\ilff}[1]{{\mathcal L}^{-1}\left[#1\right]}
\newcommand{\genm}[1]{{\mathcal M}\!\left(#1\right)}
\newcommand{\lapl}[1]{{\mathcal L}_{\mathcal M}{\left[#1\right]}}
\newcommand{\laps}[1]{{\mathcal L}_{\mathfrak{N}}{\left[#1\right]}}

\begin{document}

\title{A General Formulation of the Source Confusion Statistics
and Application to Infrared Galaxy Surveys
}
\author{
Tsutomu T. Takeuchi\altaffilmark{1,3}
\and
Takako T. Ishii\altaffilmark{2,3}
}

\altaffiltext{1}{National Astronomical Observatory of Japan,
Mitaka, Tokyo 181--8588, JAPAN; 
{\tt takeuchi@optik.mtk.nao.ac.jp}.
}

\altaffiltext{2}{
Kwasan and Hida Observatories, Kyoto University, 
Yamashina-ku, Kyoto 607--8471, JAPAN; 
{\tt ishii@kwasan.kyoto-u.ac.jp}.
}

\altaffiltext{3}{
Research Fellows of the Japan Society for the Promotion of Science
}

\begin{abstract}
Source confusion has been a long-standing problem in the astronomical history.
In the previous formulation of the confusion problem, sources are assumed to 
be distributed homogeneously on the sky.
This fundamental assumption is, however, not realistic in many applications.
In this work, by making use of the point field theory, we derive 
general analytic formulae for the confusion problems with arbitrary 
distribution and correlation functions.
As a typical example, we apply these new formulae to the source confusion of 
infrared galaxies.
We first calculate the confusion statistics for power-law galaxy number 
counts as a test case.
When the slope of differential number counts, $\gamma$, is steep, 
the confusion limits becomes much brighter and the probability 
distribution function (PDF) of the fluctuation field is strongly distorted.
Then we estimate the PDF and confusion limits based on the realistic
number count model for infrared galaxies.
The gradual flattening of the slope of the source counts makes the clustering 
effect rather mild.
Clustering effects result in an increase of the limiting flux density with 
$\sim 10\,$\%.
In this case, the peak probability of the PDF decreases up to $\sim 15$~\% 
and its tail becomes heavier.
Though the effects are relatively small, they will be strong enough to affect 
the estimation of galaxy evolution from number count or fluctuation statistics.
We also comment on future submillimeter observations.
\end{abstract}


\section{Introduction}

Astronomical source counts often suffer from a problem that multiple 
sources are located in a single beam of the observational instrument.
The obtained number counts will then be different from true ones, because 
the apparent position and flux of the sources are changed by blending of 
other, usually fainter sources.
This is called the source confusion, and the resulting measurement error is 
referred to as the confusion noise.

In its early stage, confusion problem has been investigated and formulated 
mainly by radio astronomers in relation to the so-called fluctuation 
analysis or $P(D)$ analysis 
\citep{scheuer57,hewish61,bennett62,murdoch73,condon74,condon78,wall78,wall82}.
This analysis has been immediately applied and developed further in X-ray 
\citep[e.g.,][]{scheuer74,franceschini82,barcons90,barcons92,barcons94} 
and infrared astronomy 
\citep[e.g.,][]{hacking87,franceschini89,oliver97,matsuhara00,lagache00,
miville02,friedmann03}.
Now confusion problem also becomes important in submillimeter cosmology and
high-precision astrometric missions \citep[cf.][]{hogg01}.

Cosmologists compare model predictions with observed source counts to
extract information of the evolution of galaxies and other objects
\citep[e.g.,][]{franceschini91,guiderdoni98,
takeuchi99,hirashita99,franceschini01,takeuchi01a,takeuchi01b}.
The confusion noise often prevents us from direct comparison 
between measured counts and predictions.
Especially, recent advent of infrared (IR) and submillimeter facilities may 
have stimulated the discussion on the confusion problem.
Compared with other wavelengths, we have relatively small aperture telescopes
in the IR, mainly because ground-based observations are almost
impossible, and we inevitably need cooled space telescopes.
In the submillimeter the antenna is relatively large up to 
$\sim 10\mbox{--}30\,\mbox{m}$, but the long wavelength also makes 
the single-dish surveys confusion-limited by a large diffraction.

Generally, the measurement error of the source flux blurs the true source 
counts.
This problem was originally pointed out by \citet{eddington13} and deeply 
considered in subsequent studies 
\citep[e.g.,][]{eddington40,bennett62,refsdal69,murdoch73,hogg98}, but there 
still remains some unsolved issues, especially when the error is
dominated by the confusion noise. 
Recently Monte Carlo simulation has become popular to evaluate confusion
effect \citep[e.g.,][]{bertin97,eales00,scott02}, but such an approach 
is sometimes not easy to interpret, and often time consuming.
Today, many large survey projects will be performed or completed soon, 
and a general analytical prescription for the confusion is desirable.

Fundamental assumption for the formulation of the confusion problem is 
that the sources are distributed homogeneously on the sky \citep[e.g.,][]{
scheuer57,condon74,franceschini89}.
However, this assumption is obviously not realistic in many applications, 
e.g., stars in the Galaxy, or galaxies in the Universe.
A straightforward attempt to take into account the source clustering is to 
integrate over the spatial correlation functions along the line of sight.
This approach has been adopted by several works 
\citep[e.g.,][]{barcons88,barcons98,burigana98}.
Among them, \citet{toffolatti98} presented comprehensive results closely 
related to the present issue.
Since it requires the knowledge of {\sl three-dimensional correlation 
functions}, while it is a convenient method for theoretical predictions,
observationally it is quite rare that we can obtain spatial information
at the early phase of a survey.
Hence, we need a method to evaluate the confusion {\sl only from the projected
information of source clustering}.
In this line, \citet{barcons92} has pioneered the methodology to tackle
the riddle, but his results were restricted to a few simple cases.
Since then, little analytical progress on this topic has been made up to
now.

In this work, by making extensive use of the methods of the theory of
point process, we show the general analytic formula for the confusion
problems with arbitrary distribution and correlation functions.
In Section~\ref{sec:fluctuation}, we first consider the fluctuation of 
unclustered sources, and then present the general formulae for
inhomogeneous and clustered source distributions.
Based on the new formulae, we reconsider how to treat the confusion noise in 
Section~\ref{sec:confusion}.
In Section~\ref{sec:application} we formulate higher-order clustering of 
galaxies, which will be included in the general confusion problems.
We focus on far-infrared and submillimeter galaxies.
Section~\ref{sec:conclusion} is devoted to our conclusions.
Factorial moments and cumulants are introduced in
Appendix~\ref{sec:factorial}.
In Appendix~\ref{sec:cumulants} we derive the first- and second-order
cumulants of clustered field.
We present a general expression for the rule-of-thumb to avoid 
confusion in Appendix~\ref{sec:rule_of_thumb}.
Glossary of the symbols are provided in Appendix~\ref{sec:glossary}.

\section{Statistics of the Fluctuation Field}\label{sec:fluctuation}

Statistics of fluctuation field is a fundamental tool for the confusion 
problem.
We first formulate the fluctuation of confusion noise in line with classical
works \citep[][]{scheuer57,scheuer74}, and then we translate 
it to the language of the theory of point fields \citep{daley03}.
By this method, we can straightforwardly extend our formulation to 
the confusion problem of clustered sources \citep[cf.][]{barcons92}.

\subsection{Fluctuation Generated from Sources without Clustering}

\subsubsection{Classical derivation}\label{subsubsec:classical}

We define a probability density function (hereafter  PDF) of a source having 
a flux $S' \in [S, S + d S]$, $\mathfrak{p}(S)\,d S$, i.e., 
\begin{eqnarray}
  \int_0^\infty \mathfrak{p}(S)d S = 1 \, .
\end{eqnarray}
We set $h(\vect{x})=h(\theta, \phi)$, the beam pattern (normalized to unity
at the beam center), and $s(\vect{x}) = Sh(\vect{x})$, the system response
from the intrinsic flux $S$.

The total signal response we observe, $I(\vect{x})$, is described as
\begin{eqnarray}\label{eq:total_signal}
  I(\vect{x})=\sum_{n=0}^{\infty} S_n h(\vect{x}-\vect{x}_n) \,,
\end{eqnarray}
where $S_n$ is the flux of the $n$-th source, and $\vect{x}_n$ is 
the angular position of the source.
For the following discussion we need to define a flux signal that consists
of exactly $N$ sources, $I_N(\vect{x})$,
\begin{eqnarray}\label{eq:n_signal}
  I_N(\vect{x})=\sum_{n=0}^{N} S_n h(\vect{x}-\vect{x}_n) \,.
\end{eqnarray}

Consider an ensemble of the number of sources in a unit solid angle, 
$n_\ell, \ell=1, 2, \dots$, and let the mean of $n_\ell$ is $\aven$.
Here, $\mathfrak{N}(S)\equiv \aven\mathfrak{p}(S)$ gives the familiar 
differential number count of the considered sources.
Then, from the no clustering assumption, the probability of observing exactly 
$N$ sources in a beam is given by the Poisson process
\begin{eqnarray}\label{eq:poisson}
  p_N = \frac{(\aven\obeam)^N}{N!} e^{-\aven\obeam}\, 
\end{eqnarray}
where $\obeam$ is the solid angle of the beam.

Then let us consider the PDF of the value of 
$I(\vect{x})$, $f(I)$,\footnote{
This function corresponds to $P(D)$ in radioastronomical terminology.}
\begin{eqnarray}\label{eq:pdf}
  f(I) &=& \prob{I'\in [I,I+dI)} \nonumber \\
  &=& \sum_{N=0}^{\infty} \prob{\mbox{exactly $N$ sources lie in a beam}} 
  \prob{I_N(\vect{x}) \in [I,I+dI)} \nonumber \\
  &\equiv& \sum_{N=0}^{\infty} p_N \,g_N(I) \,,
\end{eqnarray}
where $g_N(I)$ denotes the PDF of $I_N(t)$.

Under a certain regularity condition, a PDF is uniquely characterized via
its Laplace transform (LT)\footnote{
If we use $it$ instead of $t$, we obtain a characteristic function (CF).
While the CF is also quite common in physical studies, here we use LT to refer 
literatures in mathematics and statistics.}.
We define the LTs of $f(I)$ and $g_N(I)$, $\lff$ and $\lfgn$, 
respectively,
\begin{eqnarray}
  &&\lff \equiv \expc{e^{-tI}}
    =\int_{-\infty}^{\infty} e^{-tI}f(I)dI \, \label{eq:lff},\\
  &&\lfgn = \int_{-\infty}^{\infty} e^{-tI}g_N(I)dI\,,\label{eq:lfgn}
\end{eqnarray}
where $\expc{\;\cdot\;}$ represents the expectation value of a random variable.
Then we obtain
\begin{eqnarray}
  \lff = \sum_{N=0}^{\infty} p_N \lfgn\,.
\end{eqnarray}
Hence, concrete expression of $\lfgn$ is our next step to have the functional
form of $f(I)$.

We consider random variables $s_n=S_n h(\vect{x}-\vect{x}')$ and their 
summation over $n$, $I_N(\vect{x})=\sum_{n=0}^{N} s_n$.
Then the PDF of a signal $I$ with $N$ summands is represented by the
following recursive convolution:
\begin{eqnarray}
  g_{N+1}(I) &=& \int^{I}_{0} g_{N}(I-s')\, g(s') ds' \; ,
  \label{eq:dfx_conv} \\
  g_{1}(I) &=& g(I) \; .\label{eq:dfx_init}
\end{eqnarray}
Since the LT of a convolution of two functions is a normal product of 
the LTs of them, the PDF of $s$,
\begin{eqnarray}\label{eq:def_gs}
  g(s) \equiv \prob{s' \in [s,s+ds)}
\end{eqnarray}
gives
\begin{eqnarray}
  \lfgn = \lfg^N  \,,
\end{eqnarray}
where
\begin{eqnarray}\label{eq:lfgo}
  \lfg = \int_{-\infty}^{\infty} e^{-ts} g(s)ds\,.
\end{eqnarray}
Then we have
\begin{eqnarray}\label{eq:cff}
  \lff &=&
  \sum_{N=0}^{\infty} \frac{(\aven\obeam)^N e^{-\aven\obeam}}{N!} 
    \lfg^N  \nonumber\\
  &=& e^{-\aven\obeam} \sum_{N=0}^{\infty} 
    \frac{(\aven\obeam)^N \lfg^N}{N!} \nonumber\\
  &=& e^{-\aven\obeam} e^{\aven\obeam \lfg} = e^{\aven\obeam [\lfg - 1]}\, .
\end{eqnarray}

The rest we should consider is the exact form of $\lfg$.
The profile of a single source is expressed as $s(\vect{x}) = S h(\vect{x})$.
Since the LT of a random variable $s$ is a statistical average of 
$e^{-ts}$, we obtain
\begin{eqnarray}\label{eq:lfg}
  \lfg = \frac{1}{\obeam} \int_{\obeam} \int_S e^{-tS h(\vect{x})} 
    \mathfrak{p} (S) dS d\vect{x}\,.
\end{eqnarray}
Here we used a simplified symbol $\int_S \equiv \int_{S=0}^{\infty}$,
i.e., integration with respect to $S$ over the range of $[0,\infty]$.
We use this notation for some other variables in the following.

Substituting Equation~(\ref{eq:lfg}) into Equation~(\ref{eq:cff}), 
we finally obtain
\begin{eqnarray}\label{eq:pdf_intensity}
  f(I) &=& \ilff{e^{\aven\obeam[\lfg-1]}}\nonumber \\
  &=& \ilff{\exp \aven\int_{\obeam} \left[ 
    \int_S e^{-tS h(\vect{x})} \mathfrak{p} (S) dS - 1\right]d\vect{x}} , 
    \nonumber \\
\end{eqnarray}
where $\ilff{\,\cdot\,}$ represents the inverse Laplace transform.
By the formula for moments
\begin{eqnarray}\label{eq:generate_moments}
  \mu_k\equiv\expc{I^k} = (-1)^k 
     \left. \frac{d^k \lff}{d t^k} \right|_{t=0} \;,
\end{eqnarray}
and cumulants (reduced moments) 
\begin{eqnarray}\label{eq:generate_cumulants}
  \kappa_k = (-1)^k \left. \frac{d^k \ln \lff}{d t^k} 
  \right|_{t=0} \;,
\end{eqnarray}
\citep[e.g.,][]{stuart94} we obtain the following beautiful expressions for 
the cumulants of $I$ 
\citep[the simplest result of Campbell's theorem:][]{campbell09,rice44}.
\begin{eqnarray}
  \kappa_1&=&\expc{I}=
    \mu_1= \aven\int_{\obeam} \int_S Sh(\vect{x})\mathfrak{p}(S)\, dSd\vect{x}
    =\int_{\obeam} \int_S Sh(\vect{x})\mathfrak{N}(S)\, 
    dSd\vect{x} \;,\label{eq:cumulants_1}\\
  \kappa_2&=&\expc{(I - \kappa_1)^2} = 
    \int_{\obeam} \int_S S^2h(\vect{x})^2\mathfrak{N}(S)\, dSd\vect{x}\, ,
    \;\dots,
    \label{eq:cumulants_2}
\end{eqnarray}
and generally, 
\begin{eqnarray}
  \kappa_k&=&
    \int_{\obeam} \int_S S^kh(\vect{x})^k\mathfrak{N}(S)\, dSd\vect{x}\, .
    \label{eq:cumulants_k}
\end{eqnarray}

\subsubsection{Alternative derivation via point field theory}
\label{subsubsec:point_field_theory}

The position of galaxies, stars, or other point sources is expressed as
a point in the field under consideration.
A mathematical technique to treat such a filed of points is called the theory
of point process, and has a long history 
\citep[e.g.,][]{stoyan94,stoyan95,daley03}.
The theory provides us very powerful tools for the problems that we consider
here.
Thus, we translate the above heuristic derivation of $f(I)$ with the language
of point process theory.
Considering the problem along this line enables us a straightforward 
extension of our discussion to the case of clustered sources.

The fluctuation filed $I(\vect{x})$ is expressed as
\begin{eqnarray}\label{eq:measure_representation}
  I(\vect{x})&=&\sum_{n=0}^{\infty} S_n h(\vect{x}-\vect{x}_n) \nonumber \\
    &=&\int_{\mathbb{R}^2} h(\vect{x}-\vect{x}') S(\vect{x}')\,
    {\cal N}\!\left(d\vect{x}'\right)\,,
\end{eqnarray}
where ${\cal N}(A)$ of Borel sets $A \in \mathbb{R}^2$ is the so-called 
`counting measure', which represents the number of points in the set $A$,
and $S(\vect{x}')$ is a fictitious stochastic process that takes a value
$S(\vect{x}_n')=S_n$ at each point $\vect{x}_n'$
\citep{daley03}.
Here we identify the celestial sphere with a real 
plane $\mathbb{R}^2$.
We define a random measure $\genm{A}$ as
\begin{eqnarray}\label{eq:random_measure}
  \genm{A}&\equiv&\int_{A} I(\vect{x}) d\vect{x} \nonumber \\
    &=& \int_{A} \left[\int_{\mathbb{R}^2} h(\vect{x}-\vect{x}') S(\vect{x}')\,
       {\cal N}\!\left(d\vect{x}'\right)\right]d\vect{x} \nonumber \\
    &=&\sum_{\vect{x}_n \in A} S_n
       \int_{\mathbb{R}^2} h(\vect{x}-\vect{x}_n) d\vect{x} \,.
\end{eqnarray}
Here we introduce a Laplace functional $\lapl{{\cal X}}$ with respect to
the random measure ${\cal M}$,
\begin{eqnarray}\label{eq:lst}
  \lapl{{\cal X}} &\equiv& \expc{\exp\left[-\int_{\mathbb{R}^2}
    {\cal X}(\vect{x})\genm{d\vect{x}}\right]} \nonumber \\
  &=& \expc{\exp\left[-\int_{\mathbb{R}^2}
    {\cal X}(\vect{x})I(\vect{x})d\vect{x}\right]} \,.
\end{eqnarray}
It is a Laplace counterpart of the characteristic functional of a random field,
both of which are often used in particle physics, turbulence theory, and 
structure formation theories in the Universe, etc.\
\citep[e.g.,][]{vlad94,frisch95,szapudi93,matsubara95}\footnote{
In statistical and particle physics, $\lapl{{\cal X}}$ is often expressed in 
the form 
\begin{eqnarray*}
  \lapl{{\cal X}} = \int \mathfrak{D}[I]\mathfrak{F}[I]
    e^{-\int {\cal X}(\vect{x})I(\vect{x})d\vect{x}}\,,
\end{eqnarray*}
in relation to the path integral formulation 
\citep[e.g.,][]{fry84a,szapudi93,matsubara95}.
Here $\mathfrak{F}[I]$ is the assigned probability functional for 
an overall configuration of the field $I(\vect{x})$.}.
We observe
\begin{eqnarray}\label{eq:lst_integrand}
  \int_{\mathbb{R}^2}{\cal X}(\vect{x})\genm{d\vect{x}} &=& 
    \int_{\mathbb{R}^2}{\cal X} (\vect{x})
    \left[\sum_{n}S_n h(\vect{x}-\vect{x}_n)\right]d\vect{x} 
    \nonumber \\
  &=&\sum_{n}S_n \int_{\mathbb{R}^2}{\cal X} (\vect{x})
    h(\vect{x}-\vect{x}_n)d\vect{x} \,.
\end{eqnarray}
Here, 
\begin{eqnarray}\label{eq:lst_for_point}
  \expc{\exp\left[-S_n \int_{\mathbb{R}^2}{\cal X} (\vect{x})
    h(\vect{x}-\vect{x}_n)d\vect{x}\right]}
  =\laps{\int_{\mathbb{R}^2}{\cal X} (\vect{x})
    h(\vect{x}-\vect{x}_n)d\vect{x}} \equiv {\cal Z}(\vect{x}_n)\,.
\end{eqnarray}
In Equation~(\ref{eq:lst_for_point}), $\laps{\,\cdot\,}$ represents the 
Laplace-Stieltjes functional with respect to the PDF of $S_n$, 
$\mathfrak{p}(S_n)$, i.e., 
\begin{eqnarray}
  \laps{\int_{\mathbb{R}^2}{\cal X}(\vect{x})
    h(\vect{x}-\vect{x}_n)d\vect{x} }\equiv
  \int_{S}\exp\left[-S_n\int_{\mathbb{R}^2}{\cal X}(\vect{x})
    h(\vect{x}-\vect{x}_n)d\vect{x}\right] \mathfrak{p}(S_n)dS_n \,.
\end{eqnarray}

Combining Equations~(\ref{eq:lst}), (\ref{eq:lst_integrand}), and 
(\ref{eq:lst_for_point}), we have an important relation
\begin{eqnarray}\label{eq:def_pgfl}
  \lapl{{\cal X}} &=& 
    \prod_n \expc{\exp \left[-S_n \int_{\mathbb{R}^2}{\cal X} (\vect{x})
    h(\vect{x}-\vect{x}_n)d\vect{x} \right]}
    \nonumber \\
  &=& \prod_n {\cal Z}(\vect{x}_n) \nonumber \\
  &=& G \left[{\cal Z}\right] \,.
\end{eqnarray}
The last step is the definition of the probability generating functional (PGFL)
of a point field, $G\left[{\cal Z}\right]$ 
\citep[cf.][]{balian89,daley03}.
This shows an important fact: the fluctuation field blurred by a beam profile 
is expressed in terms of the PGFL of the original point field.
{}To be exact, we can describe the observed fluctuation field as an explicit
functional of the true point field.

In general, $G\left[{\cal Z}\right]$ has some useful expansions
with respect to the factorial moments, factorial cumulants, and other related
summary statistics of the point field
\citep[][]{vlad94,kerscher01,daley03}.
The most familiar statistic for astrophysical studies may be the correlation
function (or equivalently, normalized factorial cumulants).
Then, $G\left[{\cal Z}\right]$ can be expressed in the following form
\citep[for the proof, see, e.g.,][]{ma85}:
\begin{eqnarray}\label{eq:pgfl_expansion}
  \ln G\left[{\cal Y}+1\right] 
  &=&\sum^{\infty}_{k=1}\frac{1}{k!}
    \idotsint\limits_{\mathbb{R}^2\times \cdots \times\mathbb{R}^2}
    {\cal Y}(\vect{x}_1) \cdots {\cal Y}(\vect{x}_k) c_{[k]} 
    d\vect{x}_1\cdots d\vect{x}_k \nonumber \\
  &=& \sum^{\infty}_{k=1}\frac{\aven^k}{k!} 
    \idotsint\limits_{\mathbb{R}^2\times \cdots \times\mathbb{R}^2}
    {\cal Y}(\vect{x}_1) \cdots {\cal Y}(\vect{x}_k) w_{k} \,
    d\vect{x}_1\cdots d\vect{x}_k \,,
\end{eqnarray}
where $c_{[k]}$ denotes the factorial cumulant of the point field (see 
Appendix~\ref{sec:factorial}), 
and $w_k=w_k(\vect{x}_1,\cdots , \vect{x}_k)$ is the angular $k$-point
correlation function of the point sources.
This relation will be used to extend our formulation to the clustered 
point sources in \S\ref{subsec:clustering}.
For a Poisson field, by its definition, all the higher order ($k\geq 2$) 
factorial cumulants vanish, and we obtain
\begin{eqnarray}\label{eq:cum_exp_poisson}
  \ln G\left[{\cal Y}+1\right] = \aven\int_{\mathbb{R}^2}
    {\cal Y}(\vect{x})d\vect{x}\,.
\end{eqnarray}
Hence, by substituting ${\cal Z}={\cal Y}+1$ in
Equation~(\ref{eq:cum_exp_poisson}) and doing some algebra, we obtain
\begin{eqnarray}
  G\left[{\cal Z}\right] = e^{\aven\int_{\mathbb{R}^2}
    \left[{\cal Z}(\vect{x})-1\right]d\vect{x}}\,.
\end{eqnarray}

Now returning back to Equation~(\ref{eq:lst}), 
\begin{eqnarray}\label{eq:lst_poisson}
  \lapl{{\cal X}} = G[{\cal Z}]
    = e^{\aven\int_{\mathbb{R}^2}
    \left[{\cal Z}(\vect{x})-1\right]d\vect{x}}\,.
\end{eqnarray}
In order to obtain the formula for the local process in a beam area $\obeam$,
we set the test function ${\cal Z}(\vect{x})$ as
\begin{eqnarray}\label{eq:local_process}
  {\cal Z}^*(\vect{x})= 1 - \left[1-{\cal Z}(\vect{x})\right]
    \mathbb{I}_{\obeam}\,,
\end{eqnarray}
where $\mathbb{I}_A$ is an indicator function of a set $A$, 
\begin{eqnarray}\label{eq:indicator_func}
  \mathbb{I}_{A}=\begin{cases}
    1 & \vect{x}\in A\,, \\
    0 & \vect{x}\not\in A\,. \\
    \end{cases}
\end{eqnarray}
Substituting Equation~(\ref{eq:local_process}) into 
Equation~(\ref{eq:lst_poisson}) yields 
\begin{eqnarray}
  \lapl{{\cal X}} &=& e^{\aven\int_{\mathbb{R}^2}
    \left[{\cal Z}^*(\vect{x})-1\right]d\vect{x}} \nonumber \\
  &=& e^{\aven\int_{\mathbb{R}^2}
    \left[{\cal Z}(\vect{x})-1\right]\mathbb{I}_{\obeam} 
    d\vect{x}} \nonumber \\
  &=& e^{\aven\int_{\obeam}
    \left[{\cal Z}(\vect{x})-1\right]d\vect{x}} \,.
\end{eqnarray}
Recall that 
\begin{eqnarray}\label{eq:laplace_beam}
  {\cal Z}(\vect{x}_n) = \laps{\int_{\mathbb{R}^2} 
    {\cal X}(\vect{x}) h(\vect{x}-\vect{x}_n)d\vect{x}}\, ,
\end{eqnarray}
we obtain
\begin{eqnarray}
  \lapl{{\cal X}}
  &=& \exp \aven\int_{\obeam} \left\{
    \laps{\int_{\mathbb{R}^2} {\cal X}(\vect{x}') h(\vect{x}'-\vect{x})
    d\vect{x}'}-1\right\}d\vect{x} \nonumber \\
  &=& \exp \aven\int_{\obeam} \left\{\int_S 
    e^{-S\int_{\mathbb{R}^2} {\cal X}(\vect{x}') h(\vect{x}'-\vect{x}) 
    d\vect{x}'}\mathfrak{p}(S)dS-1 \right\}d\vect{x} \,.
\end{eqnarray}
By setting the test function ${\cal X}(\vect{x}')=t$, we have
\begin{eqnarray}
  \lff&=&\exp \aven\int_{\obeam} \left[\int_S 
    e^{-tSh(\vect{x}'-\vect{x})}\mathfrak{p}(S)dS-1 \right] \,d\vect{x} 
    \,.\nonumber \\
  &=&\exp \aven\int_{\obeam} \left[\int_S 
    e^{-tSh(\vect{x})}\mathfrak{p}(S)dS-1 \right]\,d\vect{x} \,.
\end{eqnarray}
Inverting the Laplace transform, we again obtain the desired result
\begin{eqnarray}
  f(I)= \ilff{\exp \aven\int_{\obeam} \left[ 
    \int_S e^{-tSh(\vect{x})} \mathfrak{p} (S) dS - 1\right]d\vect{x}} \,,
    \nonumber \\
\end{eqnarray}
which is the same result with Equation~(\ref{eq:pdf_intensity}).

\subsection{The Case of Inhomogeneous Poisson Point Field}

Sometimes we face a problem that the points distribute locally Poisson but 
the intensity $m$ is spatially inhomogeneous, i.e., depends on 
the position on the sky.
The projected stellar density distribution of the Milky Way might be 
described in this way \citep[e.g.,][]{chandrasekhar50}.

If the typical angular scale of the variation of $\aven$ is larger than the
typical beam size, we only have to divide the sky into patches with the 
variation scale and derive the PDF of the fluctuation $f(I)$ in each patch
by Equation~(\ref{eq:pdf_intensity}).

On the other hand, the variation scale of $\aven$ is comparable or smaller than
the beam size, we should properly treat the variation within a beam.
In such case $\aven$ is expressed as $\aven(\vect{x})$ at the scale of our
interest.
The intensity in a set $A$, $\bar{N}(A)$, is obtained by integrating 
$\aven(\vect{x})$ over $A$, as 
\begin{eqnarray}
  \bar{N}(A)=\int_{A}\aven(\vect{x})d\vect{x}\,,
\end{eqnarray}
\citep[][pp.650--652]{cressie93}.
In this case the joint probability for some positions are still Poisson,
and the independence still holds.
It leads to the following formula
\begin{eqnarray}
  \lff = \exp \int_{\obeam} \left[\int_S 
    e^{-tSh(\vect{x})}\mathfrak{p}(S)dS-1 \right] \aven(\vect{x})d\vect{x} \,.
\end{eqnarray}
The same as the above, the PDF $f(I)$ is obtained by inverting 
the LT.

\subsection{Fluctuation Generated from Clustered Sources}
\label{subsec:clustering}

We now turn to the case of the point source with significant clustering.
Clustering of the sources makes the confusion effect more severe, because 
the variance of the source number is larger than that of Poisson fluctuation.
The importance of clustering in the fluctuation and confusion problems has 
already pointed out as early as the end of 1960's by \citet{refsdal69}, 
but he did not provided a quantitative answer in that paper.
The fluctuation analysis of sky brightness including the effect of clustering 
was first presented by \citet{barcons92} for the study of the X-ray background.
The central tool for his analysis was characteristic functional of the 
field $I(\vect{x})$.
We use its equivalent, Laplace functional here, and derive the formula in
mathematically rigorous way, in parallel with the discussion in 
\S\ref{subsubsec:point_field_theory}.

Again we start with the fluctuation field $I(\vect{x})= \sum_{n=0}^{\infty} 
S_n h(\vect{x}-\vect{x}_n)$, but this time $\vect{x}_n$ are not distributed 
at random on the sky, but have a certain correlation with each other.
The statistical properties of the field are characterized by 
$\lapl{{\cal X}}$.
Since the derivation of Equation~(\ref{eq:def_pgfl}) does not depend on the
clustering property of the source point field, we can also apply
Equations~(\ref{eq:def_pgfl}) and (\ref{eq:pgfl_expansion}).
Since the point field has non-vanishing correlation functions for
clustered sources, Equation~(\ref{eq:pgfl_expansion}) reads
\begin{eqnarray}
  \ln G\left[{\cal Z}\right] = \sum^{\infty}_{k=1}\frac{\aven^k}{k!} 
    \idotsint\limits_{\mathbb{R}^2 \times\cdots\times\mathbb{R}^2}
    [{\cal Z}(\vect{x}_1)-1] \cdots [{\cal Z}(\vect{x}_k)-1] 
    w_{k} (\vect{x}_1,\cdots,\vect{x}_k)\,d\vect{x}_1
    \cdots d\vect{x}_k \,,
\end{eqnarray}
therefore
\begin{eqnarray}
  G\left[{\cal Z}\right]  = \exp\sum^{\infty}_{k=1}\frac{\aven^k}{k!} 
    \idotsint\limits_{\mathbb{R}^2 \times\cdots\times\mathbb{R}^2}
    [{\cal Z}(\vect{x}_1)-1] \cdots [{\cal Z}(\vect{x}_k)-1]
    w_{k} (\vect{x}_1,\cdots,\vect{x}_k)\,d\vect{x}_1\cdots 
    d\vect{x}_k \,.
\end{eqnarray}
The same as the unclustered field, we set ${\cal Z}^*=1-(1-{\cal Z})
\mathbb{I}_{\obeam}$, and obtain
\begin{eqnarray}\label{eq:pgfl_expansion_clustering}
  G\left[{\cal Z}\right] = 
    \exp\sum^{\infty}_{k=1}\frac{\aven^k}{k!}
    \idotsint\limits_{\obeam\times\cdots\times\obeam}
    [{\cal Z}(\vect{x}_1)-1] \cdots [{\cal Z}(\vect{x}_k)-1] 
    w_{k} (\vect{x}_1,\cdots,\vect{x}_k) \,d\vect{x}_1\cdots 
    d\vect{x}_k \,,
\end{eqnarray}
Substituting Equation~(\ref{eq:laplace_beam}) yields
\begin{eqnarray}
  \lapl{{\cal X}}=\exp
    \sum^{\infty}_{k=1}\frac{\aven^k}{k!} 
    \idotsint\limits_{\obeam\times\cdots\times\obeam}
    \prod_{j=1}^{k} \left\{\laps{\int_{\mathbb{R}^2}{\cal X}(\vect{x})
    h(\vect{x}-\vect{x}_j)d\vect{x}}-1\right\} 
    w_{k} (\vect{x}_1,\cdots,\vect{x}_k)
    d\vect{x}_1\cdots d\vect{x}_k \,.\nonumber \\
\end{eqnarray}
Again by letting ${\cal X}(\vect{x})=t$, we have
\begin{eqnarray}\label{eq:laplace_for_cluster}
  \lff&=&\exp\sum^{\infty}_{k=1}\frac{\aven^k}{k!}
    \idotsint\limits_{\obeam\times\cdots\times\obeam}
    \prod_{j=1}^{k} \left[\int_{S_j} 
    e^{-tS_jh(\vect{x}-\vect{x}_j)}\mathfrak{p}(S_j)dS_j -1\right]
    w_{k} (\vect{x}_1,\cdots,\vect{x}_k)\,d\vect{x}_1\cdots 
    d\vect{x}_k \nonumber \\
  &=&\exp\sum^{\infty}_{k=1}\frac{\aven^k}{k!} 
    \idotsint\limits_{\obeam\times\cdots\times\obeam} 
    \prod_{j=1}^{k} \int_{S_j} \left[
    e^{-tS_jh(\vect{x}_j)}-1\right]\mathfrak{p}(S_j)dS_j 
    w_{k} (\vect{x}_1,\cdots,\vect{x}_k)\,d\vect{x}_1\cdots 
    d\vect{x}_k \,. \nonumber \\
\end{eqnarray}
Laplace inversion gives the final general formula for the PDF of intensity
fluctuation, $f(I)$.

We obtain cumulants of the PDF of $I$ by the same procedure as 
Equations~(\ref{eq:cumulants_1})--(\ref{eq:cumulants_2}):
\begin{eqnarray}
  \kappa_1&=&\int_{\obeam}
    \int_{S} Sh(\vect{x})\mathfrak{N}(S)\,dSd\vect{x}\,,
    \label{eq:cumulants_cluster_1}\\
  \kappa_2&=&\int_{\obeam}
    \int_{S} S^2h(\vect{x})^2\mathfrak{N}(S)\,dSd\vect{x} +
    \int_{\obeam}\int_{\obeam} 
    \left[\prod_{j=1}^2\int_{S_j} S_jh(\vect{x}_j)\mathfrak{N}(S_j)\,dS_j
    \right]w_{2} \,d\vect{x}_1d\vect{x}_2 \,,
    \label{eq:cumulants_cluster_2}
\end{eqnarray}
and so on (see Appendix~\ref{sec:cumulants}).
It is a natural result that $\kappa_1$ is the same as that of unclustered
sources.
On the other hand, $\kappa_2$ involves an additional term with respect to 
$w_2=w_2(\vect{x}_1,\vect{x}_2)$, which causes the fluctuation field 
overdispersed compared to that of unclustered sources.

\section{Reconsideration of the Confusion Problem}\label{sec:confusion}

The concept of confusion noise and confusion limit is closely related to 
the fluctuation field $f(I)$.
\citet{condon74} formulated the confusion limit for a power-law source 
number counts.
His formulation was then extended for general number counts by 
\citet{franceschini89}.
\citet{takeuchi01b} have made some improvement for precise calculation by 
the method.
We should note that the confusion noise affects source counts up to 
the flux much larger than that of confusion limit \citep{murdoch73}.
We reconsider the confusion problems in this section.

\subsection{Confusion Noise}

\begin{figure*}[t]
\centering\includegraphics[width=8cm]{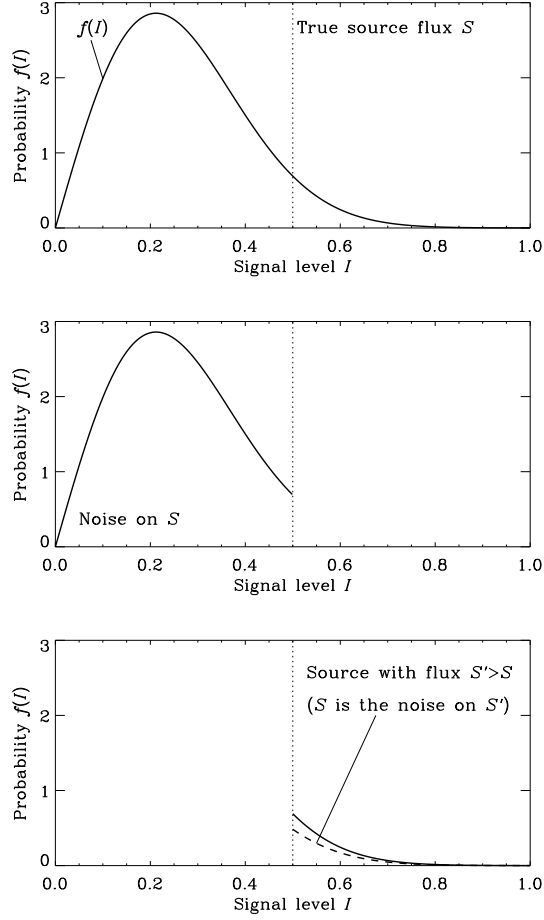}
%\epsscale{0.70}
%\plotone{f1.eps}
  \figcaption{
A schematic description of the confusion noise problem.
{\sl Top panel}: The source with flux $S$ is embedded in the fluctuation field.
{\sl Middle panel}: The source is observed with flux $S+I$ if 
the total noise intensity in the beam is $I$.
{\sl Bottom panel}: the source is observed as an additive noise of 
the brighter source with flux $S'>S$ if $I$ is dominated by one or more bright
sources $S'>S$ (dashed line).
On the other hand, if $I$ consists of some sources fainter than $S$, 
$I$ should be merely added to $S$ as a noise even if it is larger than $S$.
\label{fig:pd}}
\end{figure*}

First we review the problems related to the confusion noise.
\citet{eddington13} proposed an important concept that existence of any 
statistical noise makes the observed source counts biased upward.
Now this well-known phenomenon is called the `Eddington bias'.
He considered the case that the noise is Gaussian.
After that, a large amount of efforts has been made to treat and 
correct the bias by subsequent authors 
\citep[e.g.,][]{eddington40,bennett62,murdoch73}.
The outline of these ideas is briefly summarized in \citet{murdoch73}, in
comparison with their own method.

Problems with Gaussian noise is beautifully formulated by \citet{murdoch73}.
Their setting assumes that the dispersion of the error does not depend on
the source flux.
Even in such a simple case, \citet{murdoch73} showed that the confusion
noise can be very severe for sources with low signal-to-noise ratios.
They pointed out that when the confusion noise dominate the error, the 
total noise cannot be Gaussian, and gave a qualitative discussion on the
problem.
They claim to use a Monte Carlo approach to evaluate in that case.

However, as we will discuss in the subsequent subsections, we have an 
expression for the fluctuation field, and we can utilize the information 
for evaluating the effect of the confusion noise in general cases.
\citet{murdoch73} properly distinguish between the problems of 
{\sl confusion} and {\sl blending}: the former generally means that 
the source flux is affected by fainter sources, which cannot be detected
by the considered instrument individually, and the latter indicates that
the source flux is affected by fainter sources, which is bright enough
to be detected as a source if they were located far from the central 
brightest source.
The term `confusion limit' is related to the former phenomenon, and 
consequently, confusion limit flux is weaker than the noise flux for
bright sources.

How to treat these two effects in a unified manner?
Of course they occur in a same way, hence there is no distinction 
in principle.
Consider an ideal situation that there are no detector noise, photon noise,
read-out noise, nor any other cosmological diffuse background radiation with
structure.
Then the noise intensity distribution is the fluctuation PDF itself.

Here we consider a source with true flux $S$.
The situation is schematically described in Figure~\ref{fig:pd}.
The source is inevitably affected by other sources in a same beam.
The total noise intensity in the beam is $I$.
If $I<S$, then the source is observed with flux $S+I$ (see the middle panel of 
Figure~\ref{fig:pd}).
On the other hand, if $I$ is dominated by a bright source $S'>S$,
then the source is observed as an additive noise of the brighter 
source with flux $S'>S$ (the bottom panel of Figure~\ref{fig:pd}).
However, \citet{murdoch73} pointed out that there is a possibility that 
the signal $I$ consists of two or more sources with fluxes fainter than $S$.
The probability of such multiple source blending is obtained by the same
manner as discussed in \S\ref{sec:fluctuation}.
Again we should consider $s_j=S_jh(\vect{x}-\vect{x}_j)$, under the condition
\begin{eqnarray*}
  \sum_{j=1}^{k}s_j = I\,, 
\end{eqnarray*}
and the probability of having a noise intensity $I$ produced by $k$ sources 
is expressed by a convolution [Equation~(\ref{eq:dfx_conv})].
For $I$ to be a mere noise, there must not be sources brighter than $S$ 
in a beam.
Hence, we should derive the conditional probability of having $I$ caused by
$k$ sources under the condition $S_j<S$, $j=1,\cdots,k$, and then sum up for
all $k$s.
It may seem to be a complicated problem, but it can be simplified by 
Laplace transform.
By restricting the range of integration from 0 to $S$ in 
Equation~(\ref{eq:lfg}) we have the LT of the conditional PDF as
\begin{eqnarray}\label{eq:conditional_pdf_k}
  \mathcal{L}_{g_k}(t;S) = \left[  \frac{1}{\obeam} \int_{\obeam} 
    \int_{0}^{S} e^{-tS'h(\vect{x})} \mathfrak{p} (S') dS' 
    d\vect{x}\,\right]^k \,.
\end{eqnarray}
Then we have the LT of the desired total conditional PDF, $\tilde{f}(I;S)$, as
\begin{eqnarray}\label{eq:conditional_pdf}
  \mathcal{L}_{\tilde{f}}(t;S)  = \exp \aven\int_{\obeam} \left[ 
    \int_{0}^S e^{-tS'h(\vect{x})} \mathfrak{p} (S') dS'-1\right]d\vect{x}
    \nonumber \\
\end{eqnarray}
for unclustered point sources, and 
\begin{eqnarray}
  \mathcal{L}_{\tilde{f}}(t;S)
    =\exp\sum^{\infty}_{k=1}\frac{\aven^k}{k!}
    \idotsint\limits_{\obeam\times\cdots\times\obeam} 
    \prod_{j=1}^{k} \left[ \int_{0}^{S}
    e^{-tS_j'h(\vect{x}_j)}\mathfrak{p}(S_j')dS_j'-1\right]
    w_{k}\,d\vect{x}_1\cdots d\vect{x}_k
\end{eqnarray}
for clustered sources.

Hence, in order to obtain the confused number counts from the true ones,
we should convolve the fluctuation distribution $f(I)$ itself at $I<S$, 
and $\tilde{f}(I;S)$ at $I>S$.
This consideration naturally explains the fact that the confusion noise flux
is stronger than the confusion limit flux.
Since we have a variety of background in the image data in a real situation,
we fit and subtract from the obtained flux, the noise is not necessarily 
positive.
Thus in practice, the dependence of the confusion noise distribution 
on flux $S$ causes only a weak variation along $S$ under the existence of
other kind of noise, because of the convolution with other noise.

\subsection{Confusion limit}

We next formulate the relation between the beam size $\theta_{\rm b}$ and the
source confusion limit\footnote{Hereinafter we assume that $h(\vect{x})=
(\theta, \phi) = h(\theta)$ for simplicity. This is not an essential 
assumption for the subsequent discussions.}.
The basic procedure is to estimate the limit signal strength as an 
upper cutoff of the integration in the second-order cumulant formulae
in Equation~(\ref{eq:cumulants_2}) or Equation~(\ref{eq:cumulants_cluster_2})
by iterations.
Here we express the second-order cumulant formulae as a function of signal
strength $s$:
\begin{eqnarray}\label{eq:second_cum}
  \kappa_2&=&\int_{\obeam} \int_S S^2h(\vect{x})^2
    \mathfrak{N}(S)\,dSd\vect{x}\nonumber\\
  &=&\int_s s^2 \int_{\obeam} \mathfrak{N}\!\left[\frac{s}{h(\vect{x})}\right]
   \,\frac{d\vect{x}}{h(\vect{x})}ds \,,
\end{eqnarray}
for unclustered sources, and 
\begin{eqnarray}\label{eq:second_cum_cluster}
  \kappa_2&=&\int_{\obeam} \int_S S^2h(\vect{x})^2
    \mathfrak{N}(S)\,dSd\vect{x} %\nonumber \\
%  &&
    +
    \int_{\obeam}\int_{\obeam} 
    \int_{S_1}\int_{S_2} S_1S_2 h(\vect{x}_1)h(\vect{x}_2)
    \mathfrak{N}(S_1)\mathfrak{N}(S_2)
    w_{2}(\vect{x}_1,\vect{x}_2)dS_1dS_2 d\vect{x}_1d\vect{x}_2
    \nonumber \\
  &=&\int_{s} s^2 \int_{\obeam}
    \mathfrak{N}\left[\frac{s}{h(\vect{x})}\right]
    \frac{d\vect{x}}{h(\vect{x})}ds %\nonumber \\
%  &&
    +\int_{s_1}\int_{s_2} s_1s_2 \int_{\obeam}\int_{\obeam} 
    \mathfrak{N}\left[\frac{s_1}{h(\vect{x}_1)}\right]
    \mathfrak{N}\left[\frac{s_2}{h(\vect{x}_2)}\right]
    w_{2}\frac{d\vect{x}_1}{h(\vect{x}_1)}
    \frac{d\vect{x}_2}{h(\vect{x}_2)} ds_1ds_2 \,.
\end{eqnarray}
for clustered sources.
We utilize these formulae for deriving confusion limits in the following.

\subsubsection{Confusion Limit for power-law number counts}
\label{subsubsec:conf_lim_power}

We begin our discussion with the case that the number count is described by 
a power-law, according to \citet{condon74}\footnote{
Here $\gamma$ is the power-law exponent of differential number counts.
This is the same convention with \citet{franceschini82}.
Note that \citet{barcons92} uses the same character $\gamma$ as the exponent 
of cumulative number counts, and consequently the related expressions are 
apparently different from those in \citet{barcons92} in terms of $\gamma$.
\citet{hogg01} also use the cumulative count slope for his numerical study
of the confusion errors.}:
\begin{eqnarray}\label{eq:nc_power}
  \mathfrak{N}(S) = \alpha S^{-\gamma}\; . 
\end{eqnarray}

For unclustered sources, we obtain the confusion limit flux to a cutoff 
signal $s_{\rm c}$ from Equation~(\ref{eq:second_cum}) as
\begin{eqnarray}\label{eq:conf_lim_sigma}
  \sigma(s_{\rm c})^2 &=& \int_{0}^{s_{\rm c}} s^2 \int_{\obeam} 
    \mathfrak{N}\!\left[\frac{s}{h(\vect{x})}\right]
    \,\frac{d\vect{x}}{h(\vect{x})}ds \,,\nonumber \\
  &=&\int_{0}^{s_{\rm c}} s^2 \int_{\obeam} 
    \alpha h(\theta)^{\gamma-1} s^{-\gamma} \, \theta d\theta d\phi \nonumber\\
  &\equiv& \int_{0}^{s_{\rm c}} \alpha \Omega_{\rm eff} s^{2 - \gamma} d s 
    = \left(\frac{\alpha \Omega_{\rm eff}}{3 - \gamma}\right)
    s_{\rm c}^{3-\gamma} \, ,
\end{eqnarray}
where $\Omega_{\rm eff}$ is the so-called effective beam size, defined as
\begin{eqnarray}\label{eq:omega_eff}
  \Omega_{\rm eff} \equiv \int_{\obeam} h(\theta)^{\gamma-1}\,
    \theta d\theta d\phi \,.
\end{eqnarray}
Taking the square root of Equation~(\ref{eq:conf_lim_sigma}) and setting 
$s_{\rm c} = a \sigma$ as often used, we have 
\begin{eqnarray}\label{eq:conf_power}
  \sigma &=& 
  \left( \frac{a^{3 - \gamma}}{3 - \gamma} \right)^{1/(\gamma -1)}
  (\alpha \Omega_{\rm eff} )^{1/(\gamma -1)}. 
\end{eqnarray}
If the beam pattern is described by a Gaussian
\begin{eqnarray}\label{eq:gaussian_beam}
  h(\theta) = e^{-4\ln 2\left(\frac{\theta}{\theta_{\rm b}}\right)^2}\,,
\end{eqnarray}
where $\theta_{\rm b}$ is the FWHM of the beam, 
we get Condon's analytic formula for the $a\mbox{-}\sigma$ confusion limit,
\begin{eqnarray}\label{eq:conf_power_gauss}
  \sigma &=& 
  \left( \frac{a^{3 - \gamma}}{3 - \gamma} \right)^{1/(\gamma -1)}
  \left[ \frac{\pi\theta_{\rm b}^2\alpha}{(4 \ln 2)(\gamma - 1)} 
  \right]^{1/(\gamma -1)}.
\end{eqnarray}
It is convenient to relate these expressions and the empirical rule of thumb
for the confusion limit.
We discuss the relation between the above formulae and the empirical
rule of thumb in Appendix~\ref{sec:rule_of_thumb}.

Now we turn to the same problem for clustered sources.
For clustered sources, the confusion limit can be obtained in similar way
by substituting Equation~(\ref{eq:nc_power}) as follows:
\begin{eqnarray}\label{eq:conf_power_cluster}
  \sigma(s_{\rm c})^2&=&
    \int_{0}^{s_{\rm c}} s^2 \int_{\obeam} 
    \mathfrak{N}\!\left[\frac{s}{h(\vect{x})}\right]
    \,\frac{d\vect{x}}{h(\vect{x})}ds %\nonumber \\
%  &&
    +\int_{0}^{s_{\rm c}}\int_{0}^{s_{\rm c}} s_1s_2
    \int_{\obeam}\int_{\obeam} 
    \mathfrak{N}\left[\frac{s_1}{h(\vect{x}_1)}\right] 
    \mathfrak{N}\left[\frac{s_2}{h(\vect{x}_2)}\right]
    w_{2}\,\frac{d\vect{x}_1}{h(\vect{x}_1)}
    \frac{d\vect{x}_2}{h(\vect{x}_2)}ds_1ds_2 \nonumber \\
  &=&\int_{0}^{s_{\rm c}} s^2 \int_{\obeam} 
    \alpha h(\theta)^{\gamma-1} s^{-\gamma} \, \theta d\theta d\phi
%    \nonumber \\
%  &&
    +\int_{0}^{s_{\rm c}}\int_{0}^{s_{\rm c}} s_1s_2
    \int_{\obeam}\int_{\obeam} 
    \alpha^2 \left[h(\theta_1)h(\theta_2)\right]^{\gamma-1} (s_1 s_2)^{-\gamma}
    w_{2}\,\frac{\theta_1d\theta_1d\phi_1}{h(\theta_1)}
    \frac{\theta_2d\theta_2d\phi_2}{h(\theta_2)}ds_1ds_2 \nonumber \\
  &=&\Omega_{\rm eff}\left(\frac{\alpha}{3 - \gamma}\right)
    s_{\rm c}^{3-\gamma} + \langle{\Omega_{\rm eff}^2}\rangle
    \left( \frac{\alpha}{2-\gamma} \right)^2s_{\rm c}^{2(2-\gamma)} \,,
\end{eqnarray}
where we defined the following quantity
\begin{eqnarray}
  \langle{\Omega_{\rm eff}^2}\rangle\equiv\int_{\obeam}\int_{\obeam} 
    \left[h(\theta_1)h(\theta_2)\right]^{\gamma-1}
    w_2\,\theta_1d\theta_1d\phi_1 \theta_2d\theta_2d\phi_2 %\nonumber \\
\end{eqnarray}
In general, we cannot solve Equation~(\ref{eq:conf_power_cluster}) 
analytically, hence we should calculate it numerically.

\subsubsection{Confusion Limit for general number counts}

\citet{franceschini89} presented an iterative formula for 
the confusion limit of the general number counts.
Here we see the result for a Gaussian beam [Equation~(\ref{eq:gaussian_beam})]:
\begin{eqnarray}\label{eq:conf_general}
  \sigma^2(s_{\rm c}) &=& \int_{0}^{s_{\rm c}} s^2 \int_{\obeam} 
    \mathfrak{N}\!\left[\frac{s}{h(\vect{x})}\right]
    \,\frac{d\vect{x}}{h(\vect{x})}ds \nonumber \\
  &=& \frac{\pi \theta_{\rm b}^2}{4\ln 2} \int_{0}^{s_{\rm c}} s^2 \left[ 
    \int_{1}^{\eta_{\rm b}} \mathfrak{N} ( \eta s ) d \eta
    \right]d s \nonumber \\
  &\equiv& \frac{\pi \theta_{\rm b}^2}{4 \ln 2} J(s_{\rm c})\; ,
\end{eqnarray}
where $\eta=e^{4\ln 2 (\theta/\theta_{\rm b})}$, and $\eta_{\rm b}$ is 
the upper cutoff that corresponds to the beam area.
We again set $s_{\rm c} = a\sigma$, then we have 
\begin{eqnarray}
  \sigma = \sqrt{\frac{\pi J (a\sigma)}{4 \ln 2}}\, \theta_{\rm b}\; .
\end{eqnarray}
Thus we obtain a general relation between the beam size and the 
confusion limit as 
\begin{eqnarray}
  \theta_{\rm b} = \sqrt{\frac{4 \ln 2}{\pi J (a\sigma)}}\, \sigma \; .
\end{eqnarray}

Same as the case for power-law number counts, when source clustering takes 
place, we cannot have a simple formula, but we can still extract some 
information from the expression.
Again by setting the upper cutoff $s_{\rm c}$ in the integrals in 
Equation~(\ref{eq:second_cum_cluster}), we have
\begin{eqnarray}\label{eq:conf_general_cluster}
  \sigma^2(s_{\rm c})&=&
    \int_{0}^{s_{\rm c}} s^2 \int_{\obeam} 
    \mathfrak{N}\!\left[\frac{s}{h(\vect{x})}\right]
    \,\frac{d\vect{x}}{h(\vect{x})}ds %\nonumber \\
%  &&
    +\int_{0}^{s_{\rm c}}\int_{0}^{s_{\rm c}} s_1s_2
    \int_{\obeam}\int_{\obeam} 
    \mathfrak{N}\left[\frac{s_1}{h(\vect{x}_1)}\right] 
    \mathfrak{N}\left[\frac{s_2}{h(\vect{x}_2)}\right]
    w_2\,\frac{d\vect{x}_1}{h(\vect{x}_1)}
    \frac{d\vect{x}_2}{h(\vect{x}_2)}ds_1ds_2 \,.
\end{eqnarray}
If the beam profile is a Gaussian, and some specific functional form is 
obtained for $w_k=w_k (\vect{x}_1,\cdots,\vect{x}_k)$, similar numerical 
computation can be performed in parallel with Equation~(\ref{eq:conf_general}).

\section{Application: Galaxy Clustering and Confusion}\label{sec:application}

\subsection{Hierarchical Ansatz}

As we have discussed above, in order to treat the clustering of the sources, 
a set of prescribed angular correlation functions to some reasonable order 
are required.
Here we focus on the galaxy clustering as a typical example of the related 
issues.
Of course, we can handle any point sources as far as we have some knowledge
of their clustering.

Property of higher-order galaxy clustering still remains a matter of debate
\citep[for a comprehensive overview, see][]{peebles80}, and 
various models have been advocated in order to describe the correlation 
function of galaxies.
Generally, a hierarchy of the correlation functions continues to infinite 
order, hence we should introduce a certain closure relation to cut 
the sequence.

\begin{figure*}[t]
\centering\includegraphics[width=7cm]{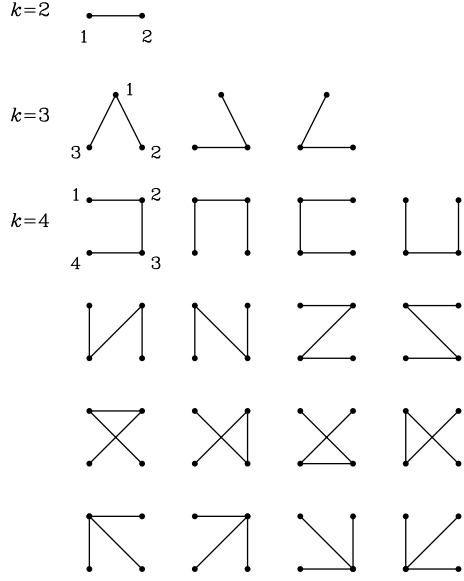}
%\epsscale{1.0}
%\plotone{f2.eps}
  \figcaption{
  The graph representation of the hierarchical expansion of the correlation
  functions.
  \label{fig:graph}}
\end{figure*}

The hierarchical {\sl Ansatz} is often employed for the correlation functions,
both from phenomenological basis \citep[e.g.,][]{balian89} and theoretical
considerations of BBGKY equations \citep[e.g.,][]{davis77,fry84b,yano97}.
The {\sl Ansatz} claims that the $k$-point correlation is described by
a product of $k-1$ two-point correlation functions.
These studies consider the three-dimensional correlations, but we can relate
them to two-dimensional ones in a simple way \citep{peebles80,szapudi93}.
We apply the following hierarchical relations for angular galaxy correlation 
on the sky \citep{meiksin92}:
\begin{eqnarray}\label{eq:hierarchical}
  w_k(\vect{x}_1,\cdots,\vect{x}_k)
    = q_k \sum_{r_j:{\rm trees}}
    w_2(\vect{x}_{r_1},\vect{x}_{r_2})\cdots
    w_2(\vect{x}_{r_{k-2}},\vect{x}_{r_{k-1}}) \,,
\end{eqnarray}
where $q_k$ is a numerical factor to determine the strength of clustering, 
and the summation $\sum_{r_j:{\rm trees}}$ is taken over all tree topologies
of graphs between points (Fig.~\ref{fig:graph}).
For the details of the expansion, see \citet{fry84a,fry84b}.

\subsection{Correlation of Infrared Galaxies}\label{sec:irgal_clustering}

Now we focus on the distribution of infrared galaxies, to have 
some insight to the confusion in the forthcoming infrared and submillimeter
galaxy surveys.
Today, many large infrared and submillimeter missions are planned or in 
progress, and their source confusion limits for galaxies have been
calculated for each facility \citep[e.g.,][]{ishii02,dole03}.
They all assumes the random distribution of galaxies on the sky.
In order to obtain the confusion limits more precisely, we must take
into account the clustering of infrared galaxies properly.

By using the density moment technique, \citet{meiksin92} estimated the 
coefficients $q_3 \mbox{--} q_8$ in Equation~(\ref{eq:hierarchical}) 
for \iras\ 1.2~Jy sample.
We use these values up to $k=4$ to approximate the clustering: $q_3=1.25$
and $q_4=2.19$.
The two-point angular correlation function is 
\begin{eqnarray}\label{eq:two_point_correlation}
  w(\vect{x}_1,\vect{x}_2) = w\left(|\vect{x}_1-\vect{x}_2|\right) =  
    w(\theta_{12}) = \left(\frac{\theta_{12}}{\theta_0}\right)^{-\beta},
\end{eqnarray}
and $\beta=0.79$, $\theta_0=0\fdg36$ for the \iras\ 1.2\,Jy sample.
We can obtain the clustering of any deeper surveys via scaling relations 
of the correlation functions with the characteristic depth of the survey, 
$d_*$ \citep{peebles80}:
\begin{eqnarray}
  w_2(\theta_{12})&=&d_*^{-1}w_2^0 (d_*\theta_{12})\,, 
    \label{eq:depth_scaling_2}\\
  w_3(\theta_{12},\theta_{23},\theta_{31})&=&
    d_*^{-2}w_3^0 (d_*\theta_{12},d_*\theta_{23},d_*\theta_{31})\,, 
    \label{eq:depth_scaling_3}\\
  w_4(\theta_{12},\theta_{23},\theta_{34},\theta_{41},\theta_{13},
    \theta_{24})&=&
    d_*^{-3}w_4^0 (d_*\theta_{12},d_*\theta_{23},d_*\theta_{34},
    d_*\theta_{41},d_*\theta_{13},d_*\theta_{24}) \,,
    \label{eq:depth_scaling_4}
\end{eqnarray}
where superscript 0 represents that it is evaluated at $S=1.2$~Jy at
$\lambda=60\,\mu$m, i.e., $w_2^0(\theta)=(\theta/0\fdg36)^{-0.79}$, and
$d_*$ is the relative characteristic depth of the survey, defined as
$d_* \equiv (1.2\,[\mbox{Jy}]/S\,[\mbox{Jy}])^{1/2}$.
We observe that the contribution of higher-order clustering rapidly decreases
with increasing depth of the survey.
This dilution of clustering is more effective for higher-order correlations.

Now we obtain the two-point correlation function $w_2(\theta)$ at an arbitrary 
flux $S$ through Equation~(\ref{eq:depth_scaling_2}) as
\begin{eqnarray}
  w_2(\theta) &=& d_*^{-(1+\beta)} w_2^0(\theta) = d_*^{-1.79} w_2^0(\theta)
    \nonumber \\
  &=& \left(\frac{1.2\,[\mbox{Jy}]}{S\,[\mbox{Jy}]}
    \right)^{-1.79/2} w_2^0(\theta) \,.
\end{eqnarray}
Consequently, from Equations~(\ref{eq:depth_scaling_3}) and 
(\ref{eq:depth_scaling_4}), we have
\begin{eqnarray}
  w_3(\theta_{12},\theta_{23},\theta_{31})
  &=& d_*^{-2(1+\beta)} w_3^0(\theta_{12},\theta_{23},\theta_{31}) 
    \nonumber \\
  &=& d_*^{-3.58} w_3^0(\theta_{12},\theta_{23},\theta_{31}) \nonumber \\
  &=& d_*^{-3.58} q_3
    \left[w_2^0(\theta_{12})+w_2^0(\theta_{23})+w_2^0(\theta_{31})\right]
    \nonumber \\
  &=& \left(\frac{1.2\,[\mbox{Jy}]}{S\,[\mbox{Jy}]}\right)^{-3.58/2} q_3 
    \left[w_2^0(\theta_{12})+w_2^0(\theta_{23})+w_2^0(\theta_{31})\right] \,,
\end{eqnarray}
and 
\begin{eqnarray}
  w_4(\theta_{12},\theta_{23},\theta_{34},\theta_{41},\theta_{13},
    \theta_{24})
  &=&
    d_*^{-3(1+\beta)}w_4^0(\theta_{12},\theta_{23},\theta_{34},\theta_{41},
    \theta_{13},\theta_{24}) \nonumber \\
  &=& d_*^{-5.37}w_4^0(\theta_{12},\theta_{23},\theta_{34},\theta_{41},
    \theta_{13},\theta_{24}) \nonumber \\
  &=& d_*^{-5.37}q_4\left[
    w_2^0(\theta_{12})+w_2^0(\theta_{23})+w_2^0(\theta_{34})
    +w_2^0(\theta_{41})+w_2^0(\theta_{13})+w_2^0(\theta_{24}) 
    \right] \nonumber \\
  &=& \left(\frac{1.2\,[\mbox{Jy}]}{S\,[\mbox{Jy}]}\right)^{-5.37/2}
%    \nonumber \\
%  &&\times 
    q_4\left[w_2^0(\theta_{12})+w_2^0(\theta_{23})+
    w_2^0(\theta_{34})
    +w_2^0(\theta_{41})+w_2^0(\theta_{13})+w_2^0(\theta_{24}) \right]\,.
    \nonumber \\
\end{eqnarray}

\bigskip

\subsection{Results}

\begin{figure*}[t]
\centering\includegraphics[width=8cm]{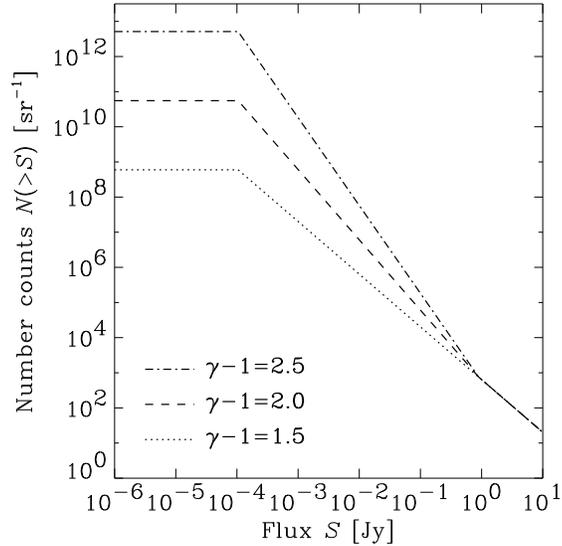}
%\plotone{f3.eps}
  \figcaption{
  The power-law number count model as a model of infrared galaxy counts.
  These model counts are the integrated ones, hence the power-law indices
  are represented by $\gamma-1=1.5$, 2.0, and 2.5.
  \label{fig:nc_power}}
\end{figure*}

\subsubsection{Power-law number counts}\label{subsubsec_power_law}

\begin{figure*}[t]
\centering\includegraphics[width=14cm]{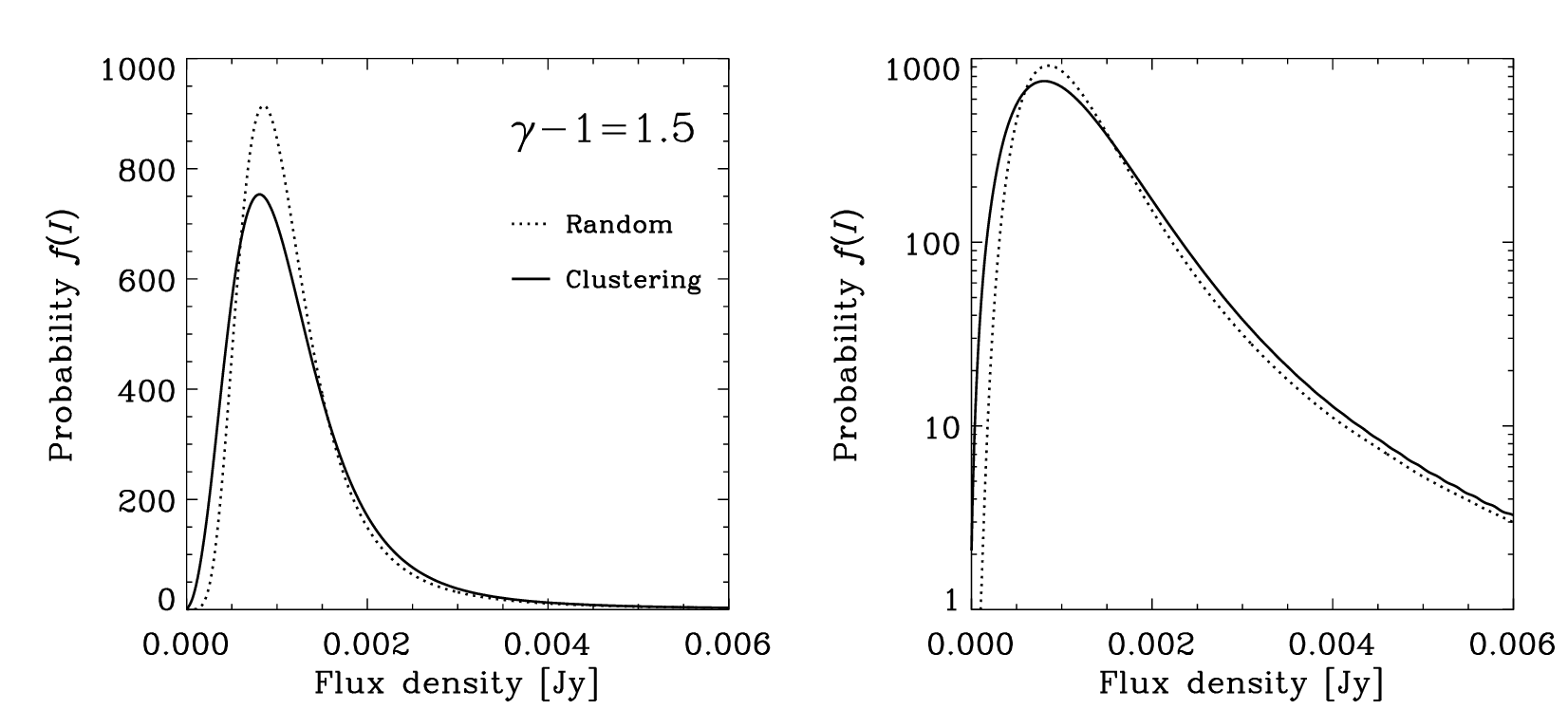}
%  \plotone{f4.eps}
  \figcaption{
  The probability density function (PDF) of the fluctuation $f(I)$ for 
  a power-law number count model with a slope index $\gamma=1.5$.
  We present the PDF in linear scale (Left panel) and in logarithmic scale
  (Right panel).
  The dotted lines show the fluctuation distribution for unclustered (randomly
  distributed) sources, whereas the solid lines represent that for the sources
  with clustering.
  \label{fig:conf_noise15}}
%\end{figure*}
%\begin{figure*}[t]
\centering\includegraphics[width=14cm]{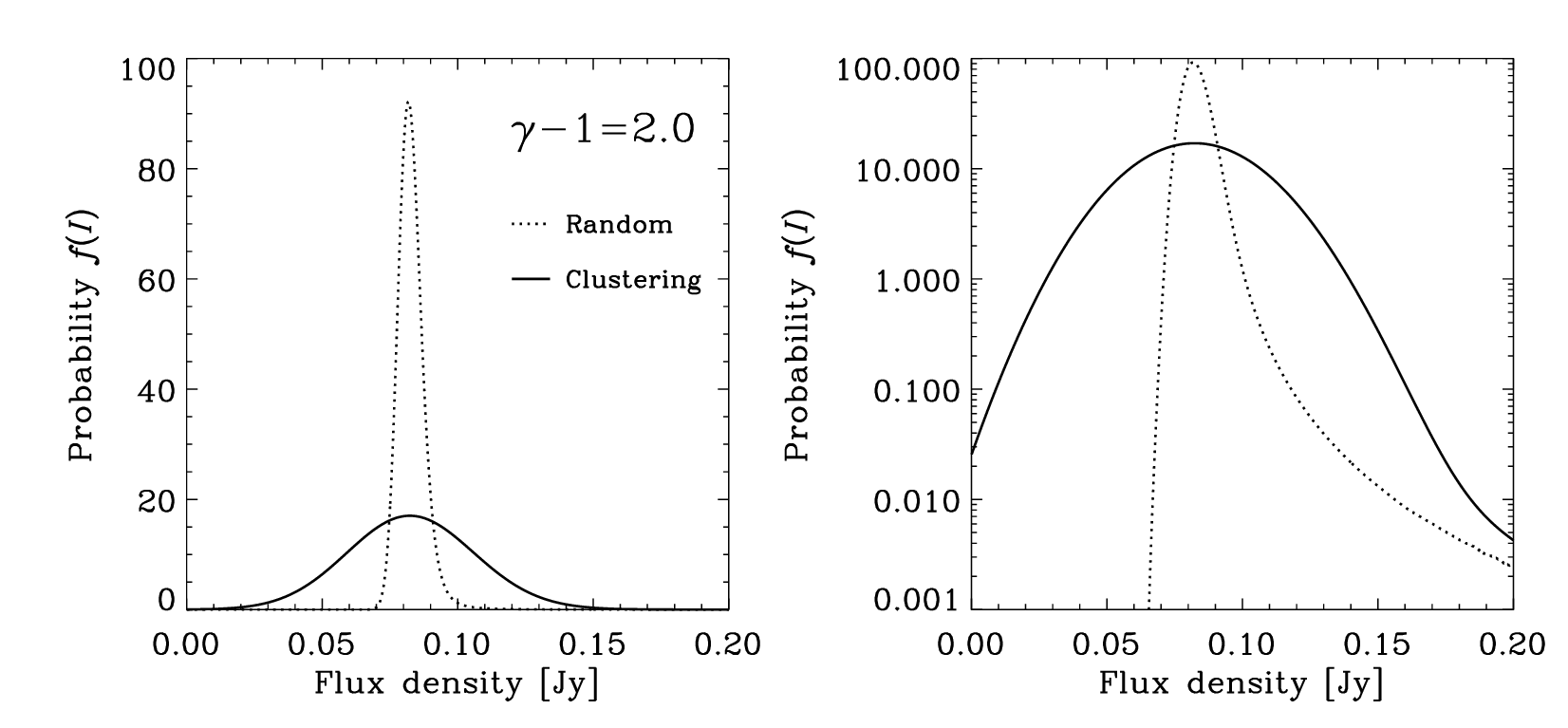}
%  \plotone{f5.eps}
  \figcaption{
  The PDF of the fluctuation for a power-law number count model with 
  a slope index $\gamma=2.0$.
  In contrast to the case of $\gamma=1.5$, the effect of clustering is
  drastic.
  \label{fig:conf_noise20}}
\end{figure*}

\begin{table}[b]
\begin{center}
\caption{The 5-$\sigma$ confusion limits for infrared galaxies 
with power-law number counts.\label{tab:conf_power_law}}
\begin{tabular}{lcc} \tableline\tableline
Index & \multicolumn{2}{c}{5-$\sigma$ confusion limits [mJy]} \\
$\gamma-1$ & Random & Clustering \\ \tableline
1.5 & 3.1 & 3.8 \\
2.0 & 22.9 & 120 \\ 
2.5 & 138 & 9332 \\ \tableline
\end{tabular}
\end{center}
\end{table}

In order to evaluate how strong the correlation structure affects the 
confusion-related quantities, we first calculate these statistics based on
power-law number count models [Equation~(\ref{eq:nc_power})],
\begin{eqnarray*}
  \mathfrak{N}(S) = \alpha S^{-\gamma}\; ,
\end{eqnarray*}
with the power-law indices $\gamma=2.5$, 3.0, and 3.5, i.e., 
the indices of the integrated counts are  $\gamma-1=1.5$, 2.0, and 2.5, 
respectively.
Here, consider a survey at $\lambda=60\,\mu$m.
The power-law number counts are normalized so that they have the same 
counts with the \iras\ QMW galaxy survey at a flux $S=0.9$~Jy 
\citep{rowan91}.
It is observationally known that the bright end of the counts is 
well approximated by Euclidean, because cosmological and evolutionary 
effects are both negligible.
Actually, \iras\ QMW galaxy counts shows a good fit to the Euclidean slope 
(slope index $\gamma=2.5$ in differential counts), and we can safely fix 
the slope of the model $\gamma$ to be 2.5 above the flux $S > 0.9$~Jy.
We estimated the confusion-related statistics for the cases with and 
without clustering.
We assume a telescope with an aperture of $70\,\mbox{cm}$ with an ideal 
Airy function as a PSF, and the detection limit of the instrument is 50~mJy.

As discussed above, the clustering of galaxies depends on the detected flux:
brighter sources have stronger clustering on the sky, and the clustering 
gradually becomes weaker toward fainter flux.
This makes the exact formulation for the statistical characteristic of 
the two-dimensional galaxy distribution prohibitively difficult
\citep[see][]{barcons92}, and unfortunately, an exact mathematical theory to 
treat this problem has not fully established yet.
In order to calculate the confusion limit with clustering, we approximate
the clustering of the whole sample galaxies evaluated at a `fiducial' flux, 
$S_{\rm fid}$, instead of the flux-dependent clustering in gradual way.
Fluctuation consists not only of the detected sources but 
also of unresolvable sources fainter than detection limit, in principle, 
toward infinitesimally faint flux.
Therefore, the fiducial flux is fainter than the point source detection limit.
We assumed that the fiducial flux is proportional to detection limit.
Based on this assumption, we calibrated the fiducial flux empirically so as to
reproduce the \iras\ confusion limit 
\citep[$\sigma \simeq 20~\mbox{mJy}$:][]{hacking87,lonsdale89,bertin97}.
We found that the relation $S_{\rm fid} = S_{\rm lim}/40$ can be used to 
evaluate the average clustering strength of the \iras\ galaxy sample.
Namely, all the galaxies with $S>S_{\rm fid}$ are approximated to have
the same correlation functions 
[$w_2(\theta)=d_{\rm fid}^{-1} w_2^0(d_{\rm fid}\theta)$, etc.]
and ignore the contribution from the sources with $S<S_{\rm fid}$.
Of course, exact treatment of this part remains to be theoretically improved.

We calculated the confusion limits from the power-law model counts.
Without clustering, the confusion limits would be 3.1~mJy, 22.9~mJy, and
138~mJy for the indices $\gamma-1=1.5, 2.0$, and 2.5, respectively.
If we take into account the effect of clustering properly, they become
3.8~mJy, 120~mJy, and 9.33~Jy, for $\gamma-1=1.5$, 2.0, and 2.5, 
respectively (see Table~\ref{tab:conf_power_law}).
Thus, the confusion-limit flux get larger, and the effect of clustering
strongly depends on the count slope.
Especially, if the cumulative count slope $\gamma-1$ exceeds $2.0$, 
its effect will be catastrophic.

We also obtained the PDF of the fluctuation intensity for infrared galaxies 
under the same assumptions.
Figure~\ref{fig:conf_noise15} shows the PDF with the integral
number count slope index $\gamma -1 = 1.5$.
The peak probability decreases by clustering.
The clustering also result in a broadening of the PDF \citep{barcons92}.
It also should noted that the probability of finding a very low intensity
at a certain position (closely related to `the void probability') increases
when clustering takes place.

The effect of clustering becomes very strong when the slope 
index approaches $\gamma-1=2.0$, as presented in Figure~\ref{fig:conf_noise20}.
In this case, we have already found that the clustering makes the confusion 
limit much shallower than that of unclustered case.
It means that the rms of the confusion noise is large.
Actually, we found that the broadening of the PDF by clustering is very
strong, and result in a large variance of the fluctuation PDF
(Fig.~\ref{fig:conf_noise20}).
In summary, the effect of clustering can be very severe if the number counts are described by {\sl a single power-law}.
The above result shows that the relative contribution of clustering to the
total fluctuation compared with Poisson component becomes larger
for large $\gamma$.

These are consistent with the results reported by \citet{toffolatti98} for
dusty galaxies at IR wavelengths.
As mentioned above, their approach is more model-oriented than the present 
work, based on the simple model of the spatial two-point correlation function 
of galaxies.
Their central aim was to estimate the power spectrum of the fluctuation in 
the submillimeter and radio background, and they considered only the two-point
correlation, i.e., second-order statistics.
Since confusion limit is the second-order quantity, hence both methods can be
used, and yield consistent estimates, though our method requires only projected
information.

Our formulation is fundamentally based on the PDF, and it is the most suitable
method of calculating it from the measured projected information 
(see \S\ref{sec:fluctuation}).
In order to extend Toffolatti et al.'s method to calculate the PDF, it is
necessary to model the bispectrum, trispectrum, and higher-order spectrum
or their equivalents \citep[for thorough formulation, see][]{matsubara03}.
It is a theoretically interesting but challenging problem, and again we see
that their method is suitable to theoretical models.

\subsubsection{Realistic number counts for infrared galaxies}

\begin{figure*}[t]
  \centering\includegraphics[width=14cm]{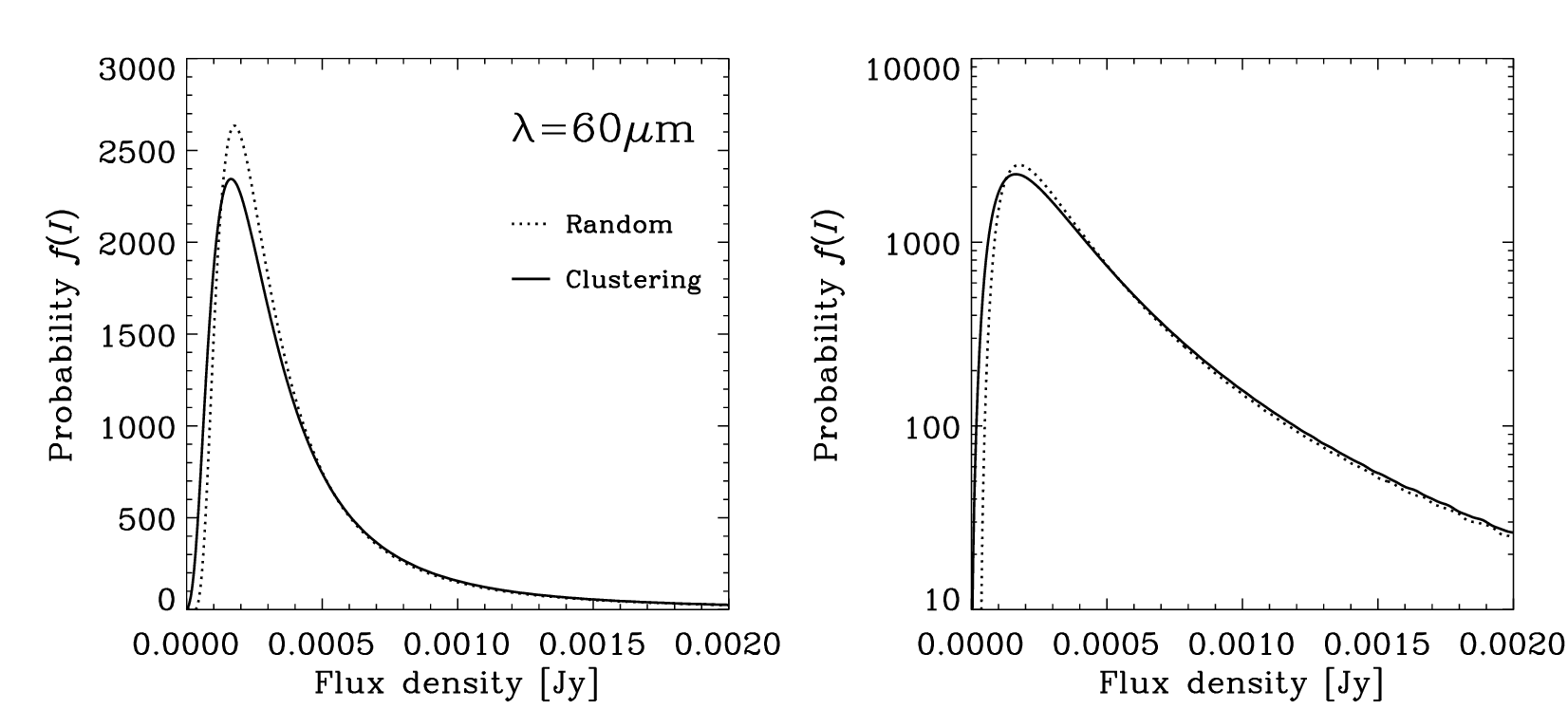}
%  \plotone{f6.eps}
  \figcaption{
  The PDF of the fluctuation $f(I)$ for a realistic number count model 
  at $60\,\mu$m based on \citet{takeuchi01a}.
  We present the PDF in linear scale (Left panel) and logarithmic scale
  (Right panel).
  The dotted lines show the PDF for unclustered (randomly distributed) 
  sources, whereas the solid lines represent the PDF for clustered sources.
  \label{fig:conf_noise060}}
%\end{figure*}
%\begin{figure*}
  \centering\includegraphics[width=14cm]{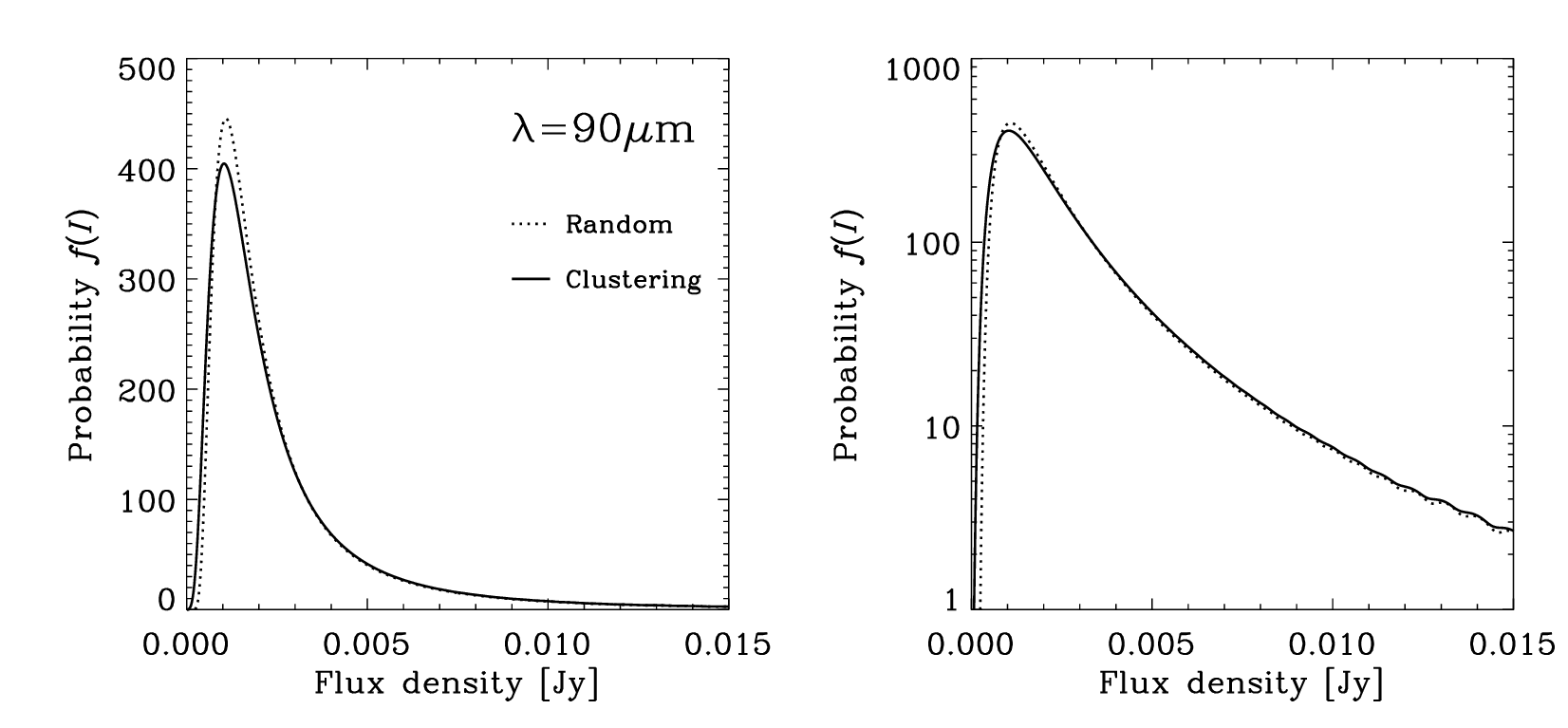}
%  \plotone{f7.eps}
  \figcaption{
  Same as Figure~\ref{fig:conf_noise060}, except that it is a PDF at 
  $90\,\mu$m. 
  \label{fig:conf_noise090}}
%\end{figure*}
%\begin{figure*}
  \centering\includegraphics[width=14cm]{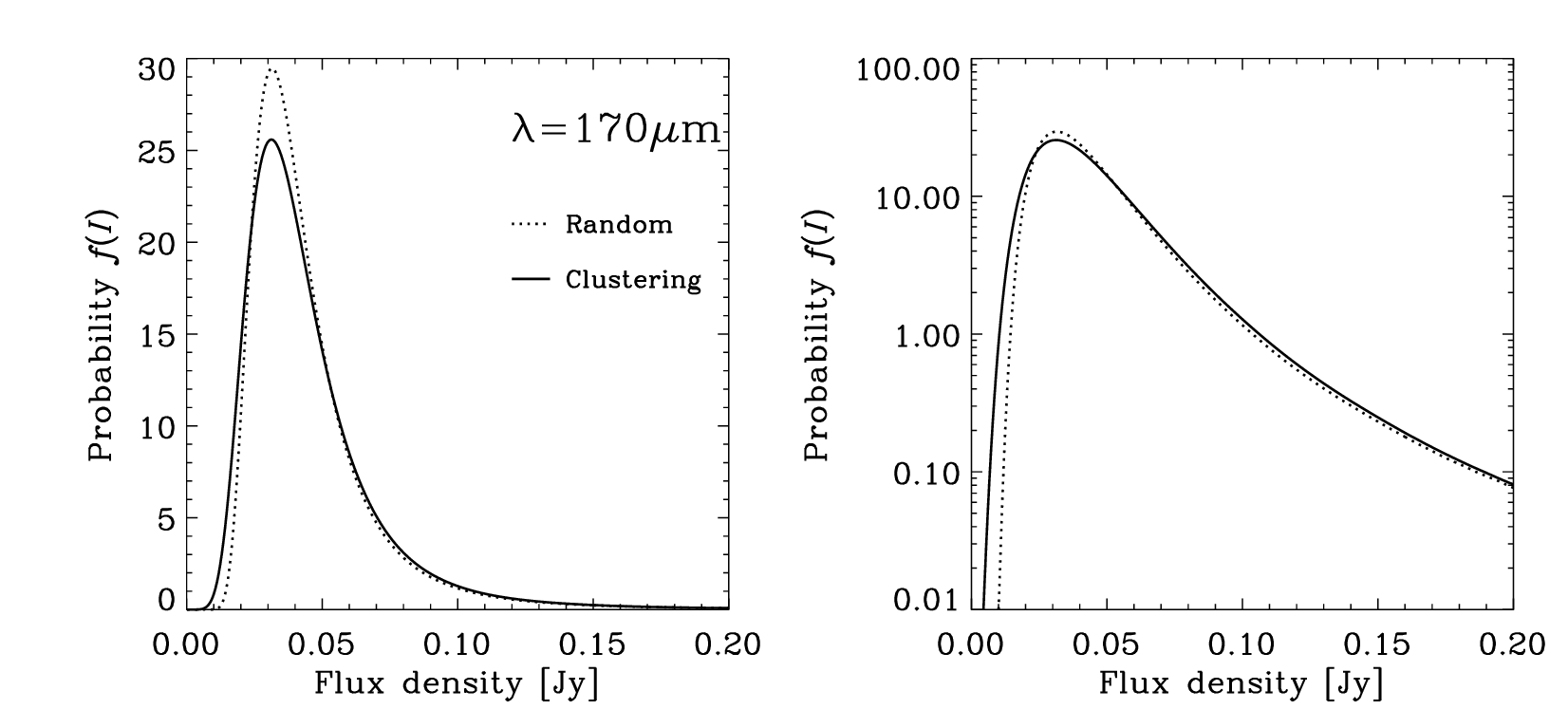}
%  \plotone{f8.eps}
  \figcaption{
  Same as Figure~\ref{fig:conf_noise060}, except that it is a PDF at 
  $170\,\mu$m.
  \label{fig:conf_noise170}}
\end{figure*}

\begin{table}
\caption{The 5-$\sigma$ confusion limits based on the model number
counts of infrared galaxies.\label{tab:conf_model}}
\begin{center}
\begin{tabular}{ccc} \tableline\tableline
Wavelength & \multicolumn{2}{c}{5-$\sigma$ confusion limits [mJy]} \\
$[\mu\mbox{m}]$ & Random & Clustering \\ \tableline
60 & 1.8 & 2.0 \\
90 & 11.4 & 12.3 \\ 
170 & 105 & 120 \\ \tableline
\end{tabular}
\end{center}
\end{table}

The observed number counts of infrared galaxies cannot be described by
a power-law function.
Deep galaxy surveys by {\sl ISO} revealed some striking features of the 
infrared galaxy counts 
\citep[see, e.g.,][]{franceschini01,pearson01,takeuchi01a,lagache03}.
Galaxies detected at 90~$\mu$m or 170~$\mu$m show a rapid increase in
number counts at $S \sim 50\mbox{--}100$~mJy, with cumulative slope 
$\gamma-1 > 2.5$.
There have never been any model that succeeded in reproducing the extremely 
steep number-count slope of {\sl ISO}\ sources perfectly, 
even though many elaborate models have been presented to date.

But such a rapid increase of infrared galaxy cannot be continued toward 
a fainter flux level, because the integrated infrared flux from galaxies
must not exceed the observational constraint of the cosmic infrared background
\citep[for the details, see,][]{takeuchi01a,franceschini01,hauser01}.
This makes a change of the slope of the observed number counts of galaxies.

We show some calculations of the confusion limits for such realistic counts
based on an empirical model of \citet{takeuchi01a}.
The parameters are taken from \citet{takeuchi02} and \citet{ishii02}.
The calculation method is the same for the case of power-law number counts, 
except that we should use Equation~(\ref{eq:conf_general_cluster}) instead
of Equation~(\ref{eq:conf_power_cluster}).

We obtained the confusion limits of 1.8~mJy for unclustered sources and
2.0~mJy for clustered sources for a model number count at $\lambda = 60\,\mu$m.
They are 11.4~mJy and 12.3~mJy at $90\,\mu$m, and 105~mJy and 120~mJy at 
$170\,\mu$m.
We summarize these limits in Table~\ref{tab:conf_model}.
Thus in the far-infrared, source clustering does not affect the confusion 
limit significantly.
This is explained as follows: Since the cosmic expansion makes the slope 
flatter toward fainter fluxes, the count slope becomes sub-Euclidean at 
the flux level we consider here, in spite of the strong galaxy evolution.
The effect of clustering is dominated by the faintest clustered sources,
hence the decrease of differential counts makes the effect rather small.

Then, let us consider the PDF of the fluctuation for realistic number counts.
We calculated the PDF for wavelengths $\lambda = 60, 90$, and $170\,\mu$m.
Figures~\ref{fig:conf_noise060}, \ref{fig:conf_noise090}, and 
\ref{fig:conf_noise170} show the PDFs of the fluctuation in these
wavelengths.
Just as seen in the power-law count models, the peak probability decreases 
up to $\sim 15$~\%, and the tail of the PDF becomes heavier.
We also find an increase of the probability of finding a small flux density.
The clustering effect may not seem very severe, because present data 
generally suffer from more serious noise sources and/or systematics.
However, we must keep this effect in mind when we estimate galaxy 
evolution from the fluctuation analysis of the data that will be obtained 
by {\sl ASTRO-F} and {\sl SIRTF}.

Remind that we assumed the constant clustering strength in this analysis.
One may expect a smaller clustering at higher redshift, but biased galaxy 
formation scenario predicts even stronger clustering for high-redshift
dusty galaxies \citep[e.g.,][]{ishii03}.
Thus, we should take care of the effect in future infrared data analysis.

\subsubsection{Comments on submillimeter galaxies}

In the submillimeter wavelengths, confusion effects are very serious
in single-dish observations \citep{blain98, takeuchi01b}.
Although some large interferometric facilities are under construction,
the confusion problem will remain an important issue that should be 
properly considered for future airborne or space submillimetric observational
projects such as BLAST (Balloon-borne Large-Aperture Sub-millimeter 
Telescope)\footnote{URL: {\tt 
http://chile1.physics.upenn.edu/blastpublic/index.shtml}.}
and {\sl Herschel Space Observatory}\footnote{URL: 
{\tt http://astro.estec.esa.nl/First/.}}.
As seen above, we need a two-point correlation function of dusty galaxies,
whose reliable observational estimate has not been available yet
\citep{scott02,almaini03,borys03}.
For the estimation of the galaxy confusion effect in future submillimeter 
large surveys \citep[see, e.g.,][and references therein]{takeuchi01b}, 
accurate estimation of the clustering properties is desired.
The measured confusion limit from SCUBA survey data was found to be 
shallower than that expected from classical formula
(D.\ H.\ Hughes 2003, private communication), hence the effect of clustering
may take place.
If we assume that the SCUBA galaxies have the same clustering property 
with \iras\ galaxies, clustering increases the confusion limit from
20~\% to 100~\%, depending on the adopted assumption for the spectral 
energy distribution of galaxies.

Most of the submillimeter galaxies are dusty vigorous starbursts 
\citep[e.g.,][]{franceschini01,takeuchi01a,takeuchi01b,totani02},
and the dominant constituent of the cosmic submillimeter background radiation.
Hence, the statistical structure of the submillimeter background carries 
crucial information of the spatial distribution of high-redshift starburst 
galaxies \citep[e.g.,][]{peacock00}.
Our analysis plays an important role as an interface between observed 
submillimeter data and sophisticated cosmological models 
\citep[e.g.,][]{magliocchetti01,perrotta03}.

\section{Conclusions}\label{sec:conclusion}

The source confusion is a long-standing problem of the astronomical history,
going back to the works of Eddington.
Fundamental assumption of the formulation is that the sources are distributed
homogeneously on the sky.
However, this assumption is not realistic in many applications, 
e.g., stars in the Galaxy, or galaxies in the Universe.
Clustering increases the confusion effect to some extent,
but it is not an easy task to formulate the effect in the confusion
problem.
By using a model for spatial correlation functions, \citet{toffolatti98}
showed a useful result for the contribution of clustering to the fluctuation
of the background at submillimeter and radio wavelengths.
Observationally, however, the method to evaluate the confusion only from
two-dimensional information of clustering is desirable.
In this line, only the work of \citet{barcons92} has made attempt to tackle 
the difficult problem and given solutions to a few simple cases.

In this work, by making extensive use of the methods related to the point
field theory, we derived general analytic formulae for 
the confusion problems with arbitrary distribution and correlation functions.
With these formulae, we can obtain the statistical properties of the
confusion noise caused by the sources with some prescribed two-dimensional 
correlation structure.

Based on the general formulae, we first calculated the confusion limits from 
the power-law galaxy number counts as a test case for the analysis of 
infrared galaxies.
We considered an infrared facility with an aperture of 70\,cm and an ideal
Airy PSF.
For galaxy clustering, we adopted the hierarchical {\sl Ansatz} to calculate
higher-order correlation functions used in our formulation.
Without clustering, the confusion limits would be 3.1mJy, 23~mJy, and
138~mJy for the indices $\gamma-1=1.5$, 2.0, and 2.5, respectively.
If we take into account the effect of clustering properly, they become
3.8~mJy, 120~mJy, and 9.9~Jy, for $\gamma-1=1.5$, 2.0, and 2.5, 
respectively.
We also obtained the probability density function (PDF) of the fluctuation
intensity, $f(I)$.
We found a broadening of the PDF by clustering, i.e., increase of the variance,
corresponding to the increase of the confusion limit flux density.
When $\gamma$ approaches to 2.0, the clustering effect becomes very severe
for a power-law number counts.

Then we estimated the PDF and confusion limits based on the realistic
number count model of \citet{takeuchi01a}, under the same hypothetical
infrared facility.
At $60\,\mu$m, we obtained confusion limits of 2.0\,mJy for 
infrared galaxies.
At $90\,\mu$m and $170\,\mu$m, the confusion limits are 12.3\,mJy and 
120\,mJy, respectively.
These estimates are not much different from the limits calculated without
clustering [1.8\,mJy ($60\,\mu$m), 11.4\,mJy ($90\,\mu$m), and 105\,mJy 
($170\,\mu$m)].
The gradual flattening of the number count slope toward the fainter flux
densities, which is observed for infrared galaxies, makes the clustering 
effect very small, even though the slope is very steep at $S \simeq
10^{-1}\mbox{--}10^{-2}\,\mbox{Jy}$.
As for the PDF at these wavelength bands, the peak probability also 
decreases up to $\sim 15$~\%, and the tail of the PDF becomes heavier.
The clustering effect may not seem very severe, because present data 
generally suffer from more serious noise sources and/or systematics.
However, we must keep this effect in mind when we estimate galaxy 
evolution from the fluctuation analysis of the data 
by {\sl ASTRO-F} and {\sl SIRTF}.

We will also apply our method to future submillimeter large surveys.
Most of the submillimeter galaxies are dusty vigorous starbursts,
and the dominant constituent of the cosmic submillimeter background radiation.
Our analysis plays an important role as an interface between observed 
submillimeter data and sophisticated complex cosmological models.

\bigskip

We sincerely thank the anonymous referee, whose comments and suggestions 
improved the clarity of this paper very much.
We offer our gratitude to Dave Hughes, Motohiro Enoki, Taihei Yano, 
Haruhiko Ueda, Seiji Ueda, Takuji Tsujimoto, Naoteru Gouda, Yasmin Friedmann, 
Hiroshi Shibai, and Hiroyuki Hirashita for their helpful and stimulating 
discussions.
We have been supported by the Japan Society of the Promotion of Science.

\appendix
\section{A.\ Factorial Moments and Cumulants}\label{sec:factorial}

We present the definition and some formulae for the important statistical
quantities, factorial moments and factorial cumulants here.
First we consider a probability generating function (PGF), $G(u)$, 
\begin{eqnarray}
  G(u) \equiv \expc{u^N},  (|u|\leq 1)\,,
\end{eqnarray}
where $N$: a nonnegative integer-valued random variable.
Here we define the factorial power. 
For any integers $n$ and $k$, the factorial powers of $n$, $n^{[k]}$, as
\begin{eqnarray}\label{eq:factorial_power}
  n^{[k]} \equiv 
    \begin{cases}
      n(n-1)\cdots(n-k+1) & k=0,\cdots,n\,, \\
      0 & k>n\,.
    \end{cases}
\end{eqnarray}
Factorial moments of $N$ are defined as
\begin{eqnarray}
  m_{[k]} \equiv \expc{N^{[k]}} = \sum_{n=0}^\infty {\sf p}_n n^{[k]}\,,
\end{eqnarray}
where
\begin{eqnarray}
  {\sf p}_n \equiv \prob{N=n}\,.
\end{eqnarray}
Then the $k$-th factorial moments, $m_{[k]}$ are obtained as the coefficients
of the following Taylor expansion,
\begin{eqnarray}\label{eq:def_factorial_moments}
  G(1+v) = 1 + \sum_{k=1}^\infty \frac{m_{[k]}v^k}{k!}\,.
\end{eqnarray}
Note that we assumed the convergence of the series.

Factorial cumulants, $c_{[k]}$, are similarly calculated by
\begin{eqnarray}\label{eq:def_factorial_cumulants}
  \ln G(1+v) = 1 + \sum_{k=1}^\infty \frac{c_{[k]}v^k}{k!}\,.
\end{eqnarray}
The first few relations are as follows:
\begin{eqnarray}
  c_{[1]} &=& m_{[1]} = \expc{N} \,,\\
  c_{[2]} &=& m_{[2]} - m_{[1]}^2 \,,\\
  c_{[3]} &=& m_{[3]} - 3m_{[2]}m_{[1]}+2m_{[1]}^3 \,.
\end{eqnarray}
For a Poisson distribution, $p_N=\aven^N e^{-\aven}/N!$, 
\begin{eqnarray}
  c_{[1]} = \aven, \quad c_{[k]} = 0 \quad (k\geq2)\,.
\end{eqnarray}
These summary statistics are extensively used in the field of structure 
formation, especially for counts-in-cells analysis 
\citep[e.g.][]{balian89,szapudi93,szapudi95,szapudi99}.
For some more details, see e.g., \citet{vlad94} and \citet{kerscher01}.

\section{B.\ Derivation of First- and Second-order Cumulants for General Field}
\label{sec:cumulants}

Here we present how to derive the first- and second-order cumulants for
the fluctuation field of clustered sources.
We start with the LT [Equation~(\ref{eq:laplace_for_cluster})]:
\begin{eqnarray*}
  \lff &=&\exp\sum^{\infty}_{k=1}\frac{\aven^k}{k!}
    \idotsint\limits_{\obeam\times\cdots\times\obeam} 
    \prod_{j=1}^{k} \int_{S_j} \left[
    e^{-tS_jh(\vect{x}_j)}-1\right]\mathfrak{p}(S_j)dS_j 
    w_{k} (\vect{x}_1,\cdots,\vect{x}_k)\,d\vect{x}_1\cdots d\vect{x}_k \,. 
\end{eqnarray*}
Then, by Equation~(\ref{eq:generate_cumulants}), we observe
\begin{eqnarray}
  \kappa_1 &=& (-1) \left. \frac{d \ln \lff}{d t} \right|_{t=0}\nonumber\\
  &=& -\left. \sum^{\infty}_{k=1}\frac{1}{k!}
    \idotsint\limits_{\obeam\times\cdots\times\obeam} 
    \frac{d}{dt} \left\{\prod_{j=1}^{k} \int_{S_j} \left[
    e^{-tS_jh(\vect{x}_j)}-1\right]\mathfrak{N}(S_j)dS_j \right\}
    w_{k} \,d\vect{x}_1\cdots d\vect{x}_k
    \right|_{t=0}\,. \label{eq:derive_kappa_1}
\end{eqnarray}
Here we define
\begin{eqnarray}
  U(\vect{x}_j;t) \equiv 
    \int_{S_j} \left[e^{-tS_jh(\vect{x}_j)}-1\right]\mathfrak{N}(S_j)dS_j \,,
\end{eqnarray}
then Equation~(\ref{eq:derive_kappa_1}) becomes
\begin{eqnarray}
  \kappa_1 = -\left. \sum^{\infty}_{k=1}\frac{1}{k!}
    \idotsint\limits_{\obeam\times\cdots\times\obeam}\sum_{\ell=1}^k
    \left[\frac{dU(\vect{x}_\ell;t)}{dt}\, 
    \underbrace{U(\vect{x}_1;t) 
    \cdots \widehat{U(\vect{x}_\ell;t)}\cdots U(\vect{x}_k;t)
    }_{(k-1)\,\text{terms}}
    \right]
    w_{k} \,d\vect{x}_1\cdots d\vect{x}_k
    \right|_{t=0} \,,
\end{eqnarray}
where $\widehat{~\cdot~}$ means that the term is omitted in the product.
We see that 
\begin{eqnarray}
  U(\vect{x}_j;0) = \int_{S_j} \left(1-1\right)\mathfrak{N}(S_j)dS_j =0 \,,
\end{eqnarray}
therefore, among the terms summed over $k=1, \cdots, \infty$, all those with 
$k \geq 2$ vanish and only $k=1$ term remains.
Thus we obtain
\begin{eqnarray}
  \kappa_1 &=& -\left. \int_{\obeam} \frac{dU(\vect{x};t)}{dt}\, d\vect{x}
    \right|_{t=0} 
    = -\left. \int_{\obeam} \int_S\left[-Sh(\vect{x}) e^{-tSh(\vect{x})}\right]
    \mathfrak{N}(S)\,dS d\vect{x}\right|_{t=0} \nonumber \\
    &=& \int_{\obeam} \int_S Sh(\vect{x}) \mathfrak{N}(S)\,dS d\vect{x} \,.
\end{eqnarray}

Next, we consider the second-order cumulant, $\kappa_2$:
\begin{eqnarray}
  \kappa_2 &=& (-1)^2 \left. \frac{d^2 \ln \lff}{d t^2} \right|_{t=0}
    \nonumber\\
  &=& \frac{d}{dt}\left. \sum^{\infty}_{k=1}\frac{1}{k!}
    \idotsint\limits_{\obeam\times\cdots\times\obeam} 
    \left\{\frac{d}{dt} \prod_{j=1}^{k} \int_{S_j} \left[
    e^{-tS_jh(\vect{x}_j)}-1\right]\mathfrak{N}(S_j)dS_j \right\}
    w_{k} \,d\vect{x}_1\cdots d\vect{x}_k
    \right|_{t=0} \nonumber \\
  &=& \frac{d}{dt}\left. \sum^{\infty}_{k=1}\frac{1}{k!}
    \idotsint\limits_{\obeam\times\cdots\times\obeam} 
    \sum_{\ell=1}^k\left[\frac{dU(\vect{x}_\ell;t)}{dt}\, 
    \underbrace{U(\vect{x}_1;t) 
    \cdots \widehat{U(\vect{x}_\ell;t)}\cdots U(\vect{x}_k;t)
    }_{(k-1)\,\text{terms}}
    \right]
    w_{k} \,d\vect{x}_1\cdots d\vect{x}_k
    \right|_{t=0} \nonumber \\ 
  &=& \left. \sum^{\infty}_{k=1}\frac{1}{k!}
    \idotsint\limits_{\obeam\times\cdots\times\obeam} 
    \sum_{\ell=1}^k\frac{d}{dt} \left[\frac{dU(\vect{x}_\ell;t)}{dt}\, 
    \prod_{j=1\cdots k}^{j\neq\ell}U(\vect{x}_j;t) 
    \right]
    w_{k} \,d\vect{x}_1\cdots d\vect{x}_k
    \right|_{t=0} \nonumber \\
  &=& \left.
    \sum^{\infty}_{k=1}\frac{1}{k!}
    \idotsint\limits_{\obeam\times\cdots\times\obeam} 
    \left[\sum_{\ell=1}^k\frac{d^2U(\vect{x}_\ell;t)}{dt^2}\, 
    \prod_{j=1\cdots k}^{j\neq\ell}U(\vect{x}_j;t) %\right.\nonumber \\
%  &&
    +\sum_{\ell, m=1\cdots k}^{\ell\neq m}
    \frac{dU(\vect{x}_\ell;t)}{dt} \frac{dU(\vect{x}_m;t)}{dt} 
    \prod_{j=1\cdots k}^{j\neq\ell,m}U(\vect{x}_j;t)
    \right] w_{k} \,d\vect{x}_1\cdots d\vect{x}_k
    \right|_{t=0} \,.\nonumber \\
    \label{eq:derive_kappa_2}
\end{eqnarray}
Among the first term in the square brackets $[\,\cdot\,]$ of 
Equation~(\ref{eq:derive_kappa_2}), all the summands with
$k\geq 2$ vanish by substituting $t=0$.
On the other hand, the second term in $[\,\cdot\,]$ exists only if $k\geq 2$.
Further, all the summands with $k \geq 3$ vanish by setting $t=0$.
Considering all the above items together, we obtain
\begin{eqnarray}
  \kappa_2 &=& \left.\int_{\obeam} \frac{d^2U(\vect{x};t)}{dt^2}\, d\vect{x}
    \right|_{t=0} %\nonumber \\
%    &+& 
    \left.
    +\frac{1}{2!}\int_{\obeam}\int_{\obeam} \left[
    \frac{dU(\vect{x}_1;t)}{dt}\frac{dU(\vect{x}_2;t)}{dt}+
    \frac{dU(\vect{x}_2;t)}{dt}\frac{dU(\vect{x}_1;t)}{dt}\right]
    w_2(\vect{x}_1,\vect{x}_2) d\vect{x}_1 d\vect{x}_2 \right|_{t=0} 
    \nonumber \\
  &=& \left. \int_{\obeam} \frac{d^2U(\vect{x};t)}{dt^2}\, d\vect{x}
    \right|_{t=0} + \left.
    \frac{2}{2!}\int_{\obeam}\int_{\obeam} 
    \frac{dU(\vect{x}_1;t)}{dt}\frac{dU(\vect{x}_2;t)}{dt}
    w_2(\vect{x}_1,\vect{x}_2) d\vect{x}_1 d\vect{x}_2 \right|_{t=0} 
    \nonumber \\
  &=& \left. \int_{\obeam} \int_S S^2h(\vect{x})^2 e^{-tSh(\vect{x})}
    \mathfrak{N}(S)\,dS d\vect{x}\right|_{t=0} \nonumber \\
  &&+ \left.\int_{\obeam}\int_{\obeam} \left[
    \int_{S_1} S_1h(\vect{x}_1) e^{-tS_1h(\vect{x}_1)}\mathfrak{N}(S_1)\,dS_1 
%    \right.\right.
%    \nonumber \\
%  &&\times \left.\left.
    \int_{S_2} S_2h(\vect{x}_2) e^{-tS_2h(\vect{x}_2)}\mathfrak{N}(S_2)\,dS_2
    \right] w_2(\vect{x}_1,\vect{x}_2)
    d\vect{x}_1 d\vect{x}_2 \right|_{t=0} \nonumber \\
  &=& \int_{\obeam} \int_S S^2h(\vect{x})^2\mathfrak{N}(S)\,dS d\vect{x}
%    \nonumber \\
%  &&
    +  
    \int_{\obeam}\int_{\obeam} \left[
    \int_{S_1} S_1h(\vect{x}_1)\mathfrak{N}(S_1)\,dS_1 
    \int_{S_2} S_2h(\vect{x}_2)\mathfrak{N}(S_2)\,dS_2 \right]
    w_2(\vect{x}_1,\vect{x}_2)
    d\vect{x}_1 d\vect{x}_2 \,.
\end{eqnarray}

\section{C.\ The Empirical Rule of Thumb for the Confusion Limit}
\label{sec:rule_of_thumb}

The `1/30 sources per beam' rule is well known to observational astronomers 
as an empirical convention to avoid confusion effects.
This problem was first theoretically addressed by \citet{franceschini82}.
Here we extend his discussion and present a numerical result for the
case of Airy beam.
According to the notation of \citet{hogg01}, we use the notation $(s/b)$ as the
number of sources per beam:
\begin{eqnarray}\label{eq:source_per_beam_s}
  (s/b)(S) \equiv \obeam\int_{S}^\infty \mathfrak{N}(S')dS' \,.
\end{eqnarray}
If we would like to detect $a$-$\sigma$ sources safely, where $\sigma$ 
is the confusion limit [Equation~(\ref{eq:conf_power})], the rule of thumb is 
expressed as 
\begin{eqnarray}
  (s/b)_{\rm C} = \obeam\int_{a\sigma}^\infty \mathfrak{N}(S)dS \,,
\end{eqnarray}
where the subscript C represents that $(s/b)$ at the confusion limit.
For power-law source counts, we can obtain a useful analytic expression for
$(s/b)_{\rm C}$ as follows:
\begin{eqnarray}\label{eq:source_per_beam}
  (s/b)_{\rm C} &=& \obeam\int_{a\sigma}^\infty \alpha S^{-\gamma} dS 
    \nonumber \\
  &=& \frac{\alpha\obeam}{\gamma-1}(a\sigma)^{1-\gamma} \nonumber \\
  &=& \frac{\alpha\obeam}{\gamma-1}a^{1-\gamma} 
    \left( \frac{a^{3 - \gamma}}{3 - \gamma} \right)^{(1-\gamma)/(\gamma -1)}
    (\alpha \Omega_{\rm eff} )^{(1-\gamma)/(\gamma -1)} \nonumber \\
  &=& \frac{\alpha\obeam}{\gamma-1}a^{1-\gamma} 
    \left( \frac{a^{3 - \gamma}}{3 - \gamma} \right)^{-1}
    (\alpha \Omega_{\rm eff} )^{-1} \nonumber \\
  &=& \frac{\obeam}{\Omega_{\rm eff}}
    \left( \frac{3 - \gamma}{\gamma-1} \right)a^{-2} \,.
\end{eqnarray}
This corresponds to the result obtained for the general case by 
\citet{franceschini82} in his Equation~(14).
For a Gaussian beam, Equation~(\ref{eq:source_per_beam}) becomes simpler.
If we set $\obeam=\pi(\epsilon\sigma_{\rm G})^2$, where $\sigma_{\rm G}
= \theta_{\rm b}/2\sqrt{2\ln 2}$ denotes the standard deviation of the 
Gaussian beam profile, we have
\begin{eqnarray}
  \frac{\obeam}{\Omega_{\rm eff}}=\frac{\epsilon^2 (\gamma-1)}{2} \,.
\end{eqnarray}
Here $\epsilon$ is a numerical factor that depends on the details of the 
source detection algorithm and beam profile.
We regard $\epsilon=2$ as a reasonable choice.
\citet{hogg01} defined $\obeam$ to be $\pi {\sigma_{\rm G}}^2$, which is 
different from our convention by a factor of $\epsilon$.
Then, we observe the following formula
\begin{eqnarray}
  (s/b)_{\rm C} = \frac{\epsilon^2 (3-\gamma)}{2a^2} \,.
\end{eqnarray}
This formula clearly shows the dependence of the confusion rule of thumb,
$(s/b)_{\rm C}$: the steeper the slope is, the more severe the confusion 
becomes.
This trend has already known to observational astronomers empirically, 
and confirmed by the comprehensive work of \citet{hogg01}.
We tabulate the rule of thumb when we take $a=5$ and $\epsilon=2$ in 
Table~\ref{tab:rule_of_thumb}.

\begin{figure*}[t]
\figurenum{9}
\centering\includegraphics[width=8cm]{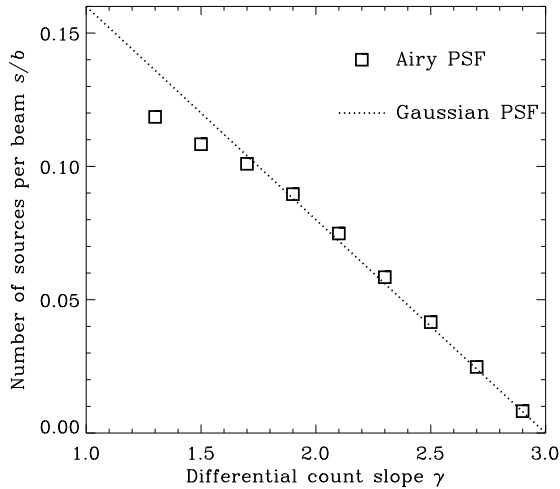}
%\plotone{f9.eps}
\figcaption{The rule of thumb for the confusion limit of 5-$\sigma$ sources.
Dotted line represents the number of sources per beam $(s/b)_{\rm C}$ 
for the confusion limit of a Gaussian beam, as a function of the slope $\gamma$
of the differential source counts. 
Open squares are $(s/b)_{\rm C}$ for an Airy beam.
These estimates are calculated for unclustered sources.
\label{fig:rule_of_thumb}
}
\end{figure*}

Generally, an ideal beam pattern of an optical instrument is often 
described by an Airy function.
We calculated the rule of thumb for an Airy beam by numerical integration.
This is shown in Figure~\ref{fig:rule_of_thumb}.
The numbers of sources per beam for Gaussian and Airy beam show an excellent
agreement at $\gamma > 1.7$.
However, we find a slight deviation from Gaussian value at 
$1.3 \la \gamma \la 1.7$ for $(s/b)_{\rm C}$ of an Airy beam. 
This may stem from the structure of the Airy function.

\begin{table}
\tablenum{3}
\begin{center}
\caption{The rule of thumb for the confusion limit of 5-$\sigma$ sources,
when the beam is Gaussian.
\label{tab:rule_of_thumb}}
\begin{tabular}{cc} \tableline\tableline
Index & Sources per beam\\
$\gamma$ & $(s/b)_{\rm C}$ \\ \tableline
0.5 & 0.20 \\
1.0 & 0.16 \\
1.5 & 0.12 \\
2.0 & 0.08 \\
2.5 & 0.04 \\
2.9 & 0.008 \\\tableline
\end{tabular}
\end{center}
\end{table}

\section{D.\ Glossary of Mathematical Symbols}\label{sec:glossary}

We tabulate the mathematical symbols used in the main text in 
Table~\ref{tab:glossary}, for the readers' convenience.
We tabulate the equations or sections by which these symbols are defined or
explained.

\begin{deluxetable}{llc}
\tablenum{4}
\tabletypesize{\footnotesize}
\tablewidth{14.5cm}
\tablecaption{Glossary of mathematical symbols.\label{tab:glossary}}
\tablehead{
\colhead{Symbol} & \colhead{Definition} & \colhead{Equation}
}
\startdata
$\prob{\mbox{Event}}$ & Probability that a certain event occurs & 
  (\ref{eq:pdf})\\
$\expc{X}$ & Expectation value of a random variable $X$ & (\ref{eq:lff})\\
$\mathbb{I}_A$ & Indicator function of a set $A$ & (\ref{eq:indicator_func})\\
$\mathbb{R}^2$ & Real plane& \S\ref{subsubsec:point_field_theory}\\
$\obeam$ & Beam area of an instrument on the sky & (\ref{eq:local_process})\\
$\Omega_{\rm eff}$ & Effective beam size, i.e., beam area weighted by 
  $h(\theta)^{\gamma-1}$ & (\ref{eq:conf_lim_sigma})\\
$\langle \Omega_{\rm eff}^2 \rangle$ & Squared beam area weighted by
  $h(\theta_1)^{\gamma-1}$, $h(\theta_2)^{\gamma-1}$, and $w_2(\theta_{12})$ & 
  (\ref{eq:conf_power_cluster})\\
$\alpha$ & Proportional constant for a power-law source number counts & 
  (\ref{eq:nc_power})\\
$\beta$ & Slope index of the power-law angular correlation function of galaxies
  & (\ref{eq:two_point_correlation})\\
$\gamma$ & Exponent of a power-law source number counts & 
  (\ref{eq:nc_power})\\
$\theta, \phi, \theta_1, \theta_2, \cdots$ & Angular coordinate or separation
  on the sky & (\ref{eq:conf_lim_sigma})\\
$\theta_{ij}$ & Angular separation between objects $i$ and $j$ &
  (\ref{eq:two_point_correlation}) \\
$\theta_{\rm b}$ & FWHM of the beam pattern & (\ref{eq:gaussian_beam})\\
$\theta_0$ & Correlation scale length of angular correlation & 
  \ref{eq:two_point_correlation}\\
$\kappa_k$ & $k$-th order cumulant of a random variable & 
  (\ref{eq:generate_cumulants})\\
$\mu_k$ & $k$-th order moment of a random variable & 
  (\ref{eq:generate_moments})\\
$\sigma$ & Confusion limit flux to a cutoff of $a\mbox{-}\sigma$ level & 
  (\ref{eq:conf_power})\\
$\sigma(s_{\rm c})$ & Confusion limit flux to a cutoff signal $s_{\rm c}$ & 
  (\ref{eq:conf_lim_sigma})\\
$\sigma_{\rm G}$ & Standard deviation of a Gaussian beam profile & 
  Appendix~\ref{sec:rule_of_thumb}\\
%$\mathfrak{D}[I]$ & Wiener (path integral) measure & 
%  \S\ref{subsubsec:point_field_theory} \\
%$\mathfrak{F}[I]$ & Assigned probability functional of $I(\vect{x})$ & 
%  \S\ref{subsubsec:point_field_theory} \\
$G[\mathcal{Z}]$ & Probability generating functional (PGFL) of a point
  field & (\ref{eq:lst_poisson})\\
$I(\vect{x})$ & Intensity of a signal at a position $\vect{x}$ & 
  (\ref{eq:total_signal})\\
$I_N(\vect{x})$ & Intensity of a signal at a position $\vect{x}$,
  consisting of exactly $N$ sources & (\ref{eq:n_signal})\\
$\lff$   & Laplace transform (LT) of $f(t)$ & (\ref{eq:lff})\\
$\mathcal{L}_{\tilde{f}}(t;S)$ & LT of $\tilde{f}$ under the condition 
  that any $S_i < S$ for any $i = 0,1,2,\cdots$ & (\ref{eq:conditional_pdf})\\
$\lfg$   & LT of $g(t)$ & (\ref{eq:lfgo})\\
$\lfgn$  & LT of $g_N(t)$ & (\ref{eq:lfgn})\\
$\mathcal{L}_{g_k}(t;S)$ & LT of $g_k$ under the condition that any $S_i < S$
  for any $i= 1, \cdots, k$ & (\ref{eq:conditional_pdf_k})\\
$\ilff{u}$ & Inverse Laplace transform of a certain function $u(t)$ & 
  (\ref{eq:pdf_intensity})\\
$\lapl{\mathcal{X}}$ & Laplace functional of a test function $\mathcal{X}$ 
  with respect to a measure $\mathcal{M}$ & (\ref{eq:lst})\\
$\laps{\mathcal{U}}$ & Laplace-Stieltjes functional of $\mathcal{U}$ 
  with respect to $\mathfrak{p}(S)$ & (\ref{eq:lst_for_point})\\
$\genm{A}$ & Random measure a random field with respect to an area $A$ & 
  (\ref{eq:random_measure})\\
$N$ & Measured number of sources in a beam & 
  (\ref{eq:n_signal})--(\ref{eq:pdf})\\
$\mathcal{N}(A)$ & Counting measure of a point field in an area $A$ & 
  (\ref{eq:measure_representation})\\
$\mathfrak{N}(S)$ & Differential source number counts with a flux
  density $S$ & \S\ref{subsubsec:classical} \\
$S$ & Detected flux density of a source  & \S\ref{subsubsec:classical} \\
$S_n$ & Detected flux density of a source with label $n$ & 
  \S\ref{subsubsec:classical} \\
$S_{\rm lim}$ & Detection limit flux for point sources &
  \S\ref{subsubsec_power_law} \\
$S_{\rm fid}$ & Fiducial flux to approximate the average correlation
  of a sample & \S\ref{subsubsec_power_law} \\
$\mathcal{X}$ & Test function, with which a value is fixed for a
  functional & (\ref{eq:lst})\\
$\mathcal{Y}$ & Test function of a functional & (\ref{eq:pgfl_expansion})\\
$\mathcal{Z}(\vect{x}_n)$ & Laplace-Stieltjes functional of
  $\int_{\mathbb{R}^2} \mathcal{X}(\vect{x})h(\vect{x}-\vect{x}_n)d\vect{x}$ 
  & (\ref{eq:lst_for_point}), (\ref{eq:def_pgfl})\\
$\mathcal{Z}^*(\vect{x}_n)$ & Test function $\mathcal{Z}(\vect{x}_n)$ 
  but locally restricted within a beam & (\ref{eq:local_process}) \\
$a$ & Flux density level measured by standard deviation of the noise & 
  \S\ref{subsubsec:conf_lim_power}\\
$c_{[k]}$ & $k$-th factorial cumulant & (\ref{eq:pgfl_expansion}),
  (\ref{eq:def_factorial_cumulants})\\
$d_*$ & Relative characteristic depth of a survey with respect to 
  $S=1.2\,\mbox{Jy}$ & 
  (\ref{eq:depth_scaling_2})--(\ref{eq:depth_scaling_4}) \\
$f(I)$ & Probability density function (PDF) of the signal intensity of
  the image & (\ref{eq:pdf})\\
$g(I)$ & PDF of $I_1$, i.e., signal produced by a single source & 
  (\ref{eq:dfx_init}), (\ref{eq:def_gs})\\
$g_N(I)$ & PDF of $I_N$, i.e., signal produced by exactly $N$ sources & 
  (\ref{eq:pdf}), (\ref{eq:dfx_conv})\\
$h(\vect{x})$ & Beam pattern of the instrument & \S\ref{subsubsec:classical}\\
$m_{[k]}$ & $k$-th factorial moment & (\ref{eq:def_factorial_moments})\\
$\aven$ & Mean of a source number density on the sky, $n_\ell$ $(\ell=1,2,
  \dots$) & \S\ref{subsubsec:classical}\\
$n^{[k]}$ & $k$-th factorial power of a certain integer $n$ & 
  (\ref{eq:factorial_power})\\
$p_N$ & Probability of obtaining a number $N$ for a Poisson random variable & 
  (\ref{eq:poisson})\\
$\mathfrak{p}(S)$ & Probability of finding a source with a flux $S$ &
  \S\ref{subsubsec:classical}\\
$q_k$ & Coefficient of the hierarchical model for the $k$-point
  correlation function & (\ref{eq:hierarchical})\\
$s(\vect{x})$ & Intensity of the signal at position $\vect{x}$, 
  $s(\vect{x}) \equiv S h(\vect{x})$ & \S\ref{subsubsec:classical}\\
$s_{\rm c}$ & Cutoff flux level used for calculating confusion limit & 
  (\ref{eq:conf_lim_sigma})\\
$s_n(\vect{x})$ & Intensity of the signal at $\vect{x}$ produced by 
  a single source at $\vect{x}_n$ & \S\ref{subsubsec:classical} \\
$(s/b)_{\rm C}$ & Number of sources per beam at the confusion limit & 
  (\ref{eq:source_per_beam})\\
$(s/b)(S)$ & Number of sources per beam to a flux limit $S$ & 
  (\ref{eq:source_per_beam_s})\\
$w_k(\vect{x}_1,\cdots,\vect{x}_k)$ & Angular $k$-point correlation
  function & (\ref{eq:pgfl_expansion}), (\ref{eq:two_point_correlation})\\
$w_k^0(\vect{x}_1,\cdots,\vect{x}_k)$ & Angular $k$-point correlation
  function evaluated at $S=1.2\,\mbox{Jy}$ & 
  (\ref{eq:depth_scaling_2})--(\ref{eq:depth_scaling_4})\\
$\vect{x}$ & Position where the intensity of the field is measured & 
  \S\ref{subsubsec:classical}\\
$\vect{x}_n$ & Position of a source with label $n$ & 
  \S\ref{subsubsec:classical}\\
\enddata
\end{deluxetable}


\begin{thebibliography}{}
\bibitem[Almaini et al.(2003)]{almaini03}
  Almaini, O., et al.\ 2003, \mnras, 338, 303

\bibitem[Barcons \& Fabian(1988)]{barcons88}
  Barcons, X., \& Fabian, A.\ C.\ 1988, \mnras, 230, 189

\bibitem[Barcons \& Fabian(1990)]{barcons90}
  Barcons, X., \& Fabian, A.\ C.\ 1990, \mnras, 243, 366

\bibitem[Barcons(1992)]{barcons92}
  Barcons, X.\ 1992, \apj, 396, 460

\bibitem[Barcons et al.(1994)]{barcons94}
  Barcons, X., Branduardi-Raymont, G., Warwick, R.\ S., Fabian, A.\ C., 
  Mason, K. O., McHardy, I., \& Rowan-Robinson, M.\ 1994, \mnras, 268, 833

\bibitem[Barcons, Fabian, \& Carrera(1998)]{barcons98}
  Barcons, X., Fabian, A.\ C., \& Carrera, F.\ J.\ 1998, \mnras, 293, 60

\bibitem[Balian \& Schaeffer(1989)]{balian89}
  Balian, R., \& Schaeffer, R.\ 1989, \aap, 220, 1

\bibitem[Bennett(1962)]{bennett62}
  Bennett, A.\ S.\ 1962, \mnras, 125, 75

\bibitem[Bertin et al.(1997)]{bertin97}
  Bertin, E., Dennefeld, M., \& Moshir, M.\ 1997, A\&A, 323, 685

\bibitem[Blain, Ivison, \& Smail(1998)]{blain98}
  Blain, A.\ W., Ivison, R.\ J., \& Smail, I.\ 1998, \mnras, 296, L29

\bibitem[Borys et al.(2003)]{borys03}
  Borys, C., Chapman, S., Halpern, M., \& Scott, D.\ 2003, \mnras, 344, 385

\bibitem[Burigana \& Popa(1998)]{burigana98}
  Burigana, C., \& Popa, L.\ 1998, \aap, 334, 420

\bibitem[Campbell(1909)]{campbell09}
  Campbell, N.\ R.\ 1909, Proc.\ Cambridge Phil.\ Soc., 15, 117

\bibitem[Chandrasekhar \& M\"unch(1950)]{chandrasekhar50}
Chandrasekhar, S., \& M\"unch, G.\ 1950, \apj, 112, 380

\bibitem[Condon(1974)]{condon74}
  Condon, J. J.\ 1974, ApJ, 188, 279

\bibitem[Condon \& Dressel(1978)]{condon78}
  Condon, J.\ J., \& Dressel, L.\ L.\ 1978, \apj, 222, 745

\bibitem[Crawford(1970)]{crawford70}
  Crawford D. F., Jauncey D. L., \& Murdoch H. S.\ 1970, ApJ, 162, 405

\bibitem[Cressie(1993)]{cressie93}
  Cressie, N.\ A.\ C.\ 1993, Statistics for Spatial Data, John Wiley \& Sons, 
  Inc., New York

\bibitem[Daley \& Vere-Jones(2003)]{daley03}
  Daley, D.\ J., \& Vere-Jones, D.\ 2003, An Introduction to 
  the Theory of Point Processes, Volume I: Elementary Theory and Methods,
  2nd ed. (New York: Springer)

\bibitem[Davis \& Peebles(1977)]{davis77}
  Davis, M., \& Peebles, P.\ J.\ E.\ 1977, \apjs, 34, 425

\bibitem[Dole, Lagache, \& Puget(2003)]{dole03}
  Dole, H., Lagache, G., \& Puget, J.-L.\ 2003, \apj, 585, 617

\bibitem[Eales et al.(2000)]{eales00} 
  Eales, S., Lilly, S., Webb, T., Dunne, L., Gear, W., Clements, D., \& 
  Yun, M.\ 2000, AJ, 120, 2244

\bibitem[Eddington(1913)]{eddington13}
  Eddington, A. S.\ 1913, \mnras, 73, 359

\bibitem[Eddington(1940)]{eddington40}
  Eddington, A. S.\ 1940, \mnras, 100, 354

\bibitem[Franceschini(1982)]{franceschini82}
  Franceschini, A.\ 1982, \apss, 86, 3

\bibitem[Franceschini et al.(1989)]{franceschini89}
  Franceschini, A., Toffolatti, L., Danese, L., \& de Zotti, G.\ 1989, 
  ApJ, 344, 35

\bibitem[Franceschini et al.(1991)]{franceschini91}
  Franceschini, A., Toffolatti, L., Mazzei, P., Danese, L., \& 
  de Zotti, G.\ 1991, A\&AS 89, 285

\bibitem[Franceschini et al.(2001)]{franceschini01}
  Franceschini, A., Aussel, H., Cesarsky, C. J., Elbaz, D., \& Fadda, D.\ 
  2001, \aap, 378, 1

\bibitem[Friedmann \& Bouchet(2003)]{friedmann03}
  Friedmann, Y., \& Bouchet, F.\ 2003, \mnras, in press (astro-ph/0310785)

\bibitem[Frisch(1995)]{frisch95}
  Frisch, U.\ 1995, Turbulence (Cambridge: Cambridge Univ.\ Press)

\bibitem[Fry(1984a)]{fry84a}
  Fry, J.\ N.\ 1984a, \apj, 277, L5

\bibitem[Fry(1984b)]{fry84b}
  Fry, J.\ N.\ 1984b, \apj, 279, 499

\bibitem[Guiderdoni et al.(1998)]{guiderdoni98}
  Guiderdoni, B., Hivon, E., Bouchet, F.\ R., \& Maffei, B.\ 1998, 
  \mnras, 295, 877

\bibitem[Hacking \& Houck(1987)]{hacking87}
  Hacking, P., \& Houck, J.\ R.\ 1987, \apjs, 63, 311

\bibitem[Hauser \& Dwek(2001)]{hauser01}
  Hauser, M.\ G.\ \& Dwek, E.\ 2001, \araa, 39, 249

%\bibitem[Herranz, Kuruo\u{g}lu, \& Toffolatti(2003)]{herranz03}
%  Herranz, D., Kuruo\u{g}lu, E.\ E., \& Toffolatti, L.\ 2003, \mnras, 
%  submitted (astro-ph/0307114)

\bibitem[Hewish(1961)]{hewish61}
  Hewish, A.\ 1961, \mnras, 123, 167

\bibitem[Hirashita et al.(1999)]{hirashita99}
  Hirashita, H., Takeuchi, T. T., Shibai, H., \& Ohta, K.\ 1999, PASJ, 51, 81

\bibitem[Hogg(2001)]{hogg01}
  Hogg, D. W.\ 2001, \aj, 121, 1207

\bibitem[Hogg \& Turner(1998)]{hogg98}
  Hogg, D. W., \& Turner, E. L.\ 1998, PASP, 110, 727

\bibitem[Ishii, Takeuchi, \& Sohn(2002)]{ishii02}
  Ishii, T. T., Takeuchi, T. T., \& Sohn, J.-J.\ 2002, in Infrared and 
  Submillimeter Space Astronomy, ed.\ M.\ Giard, J.\ P.\ Bernard, 
  A.\ Klotz, \& I.\ Ristrorcelli (Les Ulis: EDP Sciences), 369

\bibitem[Ishii, Takeuchi, \& Enoki(2003)]{ishii03}
  Ishii, T. T., Takeuchi, T. T., \& Enoki, M.\ 2003, in 
  Multiwavelength Cosmology, ed.\ M.\ Plionis \& I.\ Georgantopoulos
  (Dordrecht: Kluwer), in press

\bibitem[Kerscher(2001)]{kerscher01}
  Kerscher, M.\ 2001, \pre, 64, 56109

\bibitem[Lagache \& Puget(2000)]{lagache00}
  Lagache, G., \& Puget, J.-L.\ 2000, A\&A, 355, 17

\bibitem[Lagache, Dole \& Puget(2003)]{lagache03}
  Lagache, G., Dole, H., \& Puget, J.-L.\ 2003, \mnras, 338, 555

\bibitem[Lonsdale \& Hacking(1989)]{lonsdale89}
  Lonsdale, C.\ J., \& Hacking, P.\ B.\ 1989, \apj, 339, 712

\bibitem[Ma(1985)]{ma85}
  Ma, S.-K.\ 1985, Statistical Mechanics (Philadelphia: World Scientific)

\bibitem[Magliocchetti et al.(2001)]{magliocchetti01}
  Magliocchetti, M., Moscardini, L., Panuzzo, P., Granato, G.\ L., 
  De Zotti, G., \& Danese, L.\ 2001, \mnras, 325, 1553

\bibitem[Matsubara(1995)]{matsubara95}
  Matsubara, T.\ 1995, \apjs, 101, 1

\bibitem[Matsubara(2003)]{matsubara03}
  Matsubara, T.\ 2003, \apj, 584, 1

\bibitem[Matsuhara et al.(2000)]{matsuhara00}
  Matsuhara, H., et al.\ 2000, A\&A, 361, 407

\bibitem[Meiksin, Szapudi, \& Szalay(1992)]{meiksin92}
  Meiksin, A., Szapudi, I., \& Szalay, A.\ 1992, \apj, 394, 87

\bibitem[Miville-Desch\^enes, Lagache, \& Puget(2002)]{miville02}
  Miville-Desch\^enes, M.-A., Lagache, G., \& Puget, J.-L.\ 2002, 
  \aap, 393, 749

\bibitem[Murdoch et al.(1973)]{murdoch73}
  Murdoch H. S., Crawford D. F., \& Jauncey D. L.\ 1973, ApJ, 183, 1

\bibitem[Oliver et al.(1997)]{oliver97}
  Oliver, S., et al.\ 1997, \mnras, 
  289, 471

\bibitem[Peacock et al.(2000)]{peacock00}
  Peacock, J.\ A., et al.\ 2000, \mnras, 318, 535

\bibitem[Pearson(2001)]{pearson01}
  Pearson, C. P.\ 2001, \mnras, 325, 1511

\bibitem[Peebles(1980)]{peebles80}
  Peebles, P.\ J.\ E.\ 1980, Large Scale Structure of the Universe 
  (Princeton: Princeton Univ.\ Press)

\bibitem[Perrotta et al.(2003)]{perrotta03}
  Perrotta, F., Magliocchetti, M., Baccigalupi, C.,
  Bartelmann, M., De Zotti, G., Granato, G.\ L., Silva, L., \& 
  Danese, L.\ 2003, \mnras, 338, 623

\bibitem[Refsdal(1969)]{refsdal69}
  Refsdal, S.\ 1969, ApJ, 155, 373

\bibitem[Rice(1944)]{rice44}
  Rice, S. O.\ 1944, Bell System Tech.\ J., 23, 282 [reprinted in Selected 
  Papers on Noise and Stochastic Processes, 2003, ed., N.\ Wax, 
  (New York: Dover)]

\bibitem[Rowan-Robinson et al.(1991)]{rowan91}
  Rowan-Robinson, M., Saunders, W., Lawrence, A., \& Leech, K.\ 1991, 
  \mnras, 253, 485 

\bibitem[Scheuer(1957)]{scheuer57}
  Scheuer, P. A. G. 1957, Proc. Cambridge Phil. Soc., 53, 764

\bibitem[Scheuer(1974)]{scheuer74}
  Scheuer, P.\ A.\ G.\ 1974, \mnras, 166, 329

\bibitem[Scott et al.(2002)]{scott02}
  Scott, S. E., et al.\ 2002, \mnras, 331, 817

\bibitem[Stoyan, Kendall, \& Mecke(1995)]{stoyan95}
  Stoyan, D., Kendall, W.\ S., \& Mecke, J.\ 1995, Stochastic Geometry and 
  its Applications, 2nd ed. (Chichester: John Wiley \& Sons)

\bibitem[Stoyan \& Stoyan(1994)]{stoyan94}
  Stoyan, D., \& Stoyan, H.\ 1994, Fractals, Random Shapes and Point Fields
  (Chichester: John Wiley \& Sons)

\bibitem[Stuart \& Ord(1994)]{stuart94} 
  Stuart, A., \&  Ord, K.\ 1994, Kendall's Advanced Theory 
  of Statistics, 6th ed.\ Vol.\ 1, Distribution Theory (London: Arnold)

\bibitem[Szapudi \& Szalay(1993)]{szapudi93}
  Szapudi, I., \& Szalay, A.\ 1993, \apj, 408, 43

\bibitem[Szapudi et al.(1995)]{szapudi95}
  Szapudi, I., Dalton, G.\ B., Efstathiou, G., \& Szalay, A.\ 
  1995, \apj, 444, 520

\bibitem[Szapudi et al.(1999)]{szapudi99}
  Szapudi, I., Colombi, S., \& Bernardeau, F.\ 1999, \mnras, 310,428

\bibitem[Takeuchi et al.(1999)]{takeuchi99}
  Takeuchi, T. T., Hirashita, H., Ohta, K., Hattori, T. G., Ishii, T. T., \& 
  Shibai, H.\ 1999, PASP, 111, 288

\bibitem[Takeuchi et al.(2001a)]{takeuchi01a}
  Takeuchi, T. T., Ishii, T. T., Hirashita, H., Yoshikawa, K, 
  Matsuhara, H., Kawara, K., \& Okuda, H.\ 2001a, PASJ, 53, 37

\bibitem[Takeuchi et al.(2001b)]{takeuchi01b}
  Takeuchi, T. T., Kawabe, R., Kohno, K., Nakanishi, K., Ishii, T. T., 
  Hirashita, H., \& Yoshikawa, K.\ 2001b, PASP, 113, 586

\bibitem[Takeuchi et al.(2002)]{takeuchi02}
  Takeuchi, T.\ T., Shibai, H., \& Ishii, T.\ T.\ 2002, 
  Adv. Sp. Res., 30, 2021

\bibitem[Toffolatti et al.(1998)]{toffolatti98}
  Toffolatti, L., Arg\"{u}eso G\'{o}mez, F., De Zotti, G., Mazzei, P., 
  Danese, L., \& Burigana, C.\ 1998, \mnras, 297, 117

\bibitem[Totani \& Takeuchi(2002)]{totani02}
 Totani, T., \& Takeuchi, T.\ T.\ 2002, ApJ, 570, 470

\bibitem[Vlad et al.(1994)]{vlad94}
  Vlad, M.\ O, Mackey, M.\ C., \& Ross, J.\ 1994, \pre, 50, 798

\bibitem[Wall(1978)]{wall78}
  Wall, J.\ V.\ 1978, \mnras, 182, 381

\bibitem[Wall et al.(1982)]{wall82}
  Wall, J.\ V., Scheuer, P.\ A.\ G., Pauliny-Toth, I.\ I.\ K., \& Witzel, A.\
  1982, \mnras, 198, 221

\bibitem[Yano \& Gouda(1997)]{yano97}
  Yano, T., \& Gouda, N.\ 1997, \apj, 487, 473
\end{thebibliography}
\end{document}